\documentclass[lettersize,journal]{IEEEtran}
\usepackage{amsmath,amsfonts}
\usepackage{algorithmic}
\usepackage{algorithm}
\usepackage{array}
\usepackage[caption=false,font=normalsize,labelfont=sf,textfont=sf]{subfig}
\usepackage{textcomp}
\usepackage{stfloats}
\usepackage{url}
\usepackage{verbatim}
\usepackage{graphicx}
\usepackage{cite}
\usepackage{soul}
\usepackage{color}
\usepackage{amsmath,amssymb,amsfonts}
\usepackage{textcomp}
\usepackage{xcolor}
\usepackage{threeparttable}

\begin{document}

\title{HASTILY: \underline{Ha}rdware-\underline{S}oftware Co-Design for Accelerating \underline{T}ransformer \underline{I}nference \underline{L}everaging Compute-in-Memor\underline{y}}


\author{Dong Eun Kim, Tanvi Sharma, and Kaushik Roy,~\IEEEmembership{Member,~IEEE,}
\thanks{Dong Eun Kim, Tanvi Sharma, and Kaushik Roy are with the Department of Electrical and Computer Engineering, Purdue University, West Lafayette, IN, 47907.}
}



\maketitle

\begin{abstract}
Transformers have become the backbone of neural network architecture for most machine learning applications. Their widespread use has resulted in multiple efforts on accelerating attention, the basic building block of transformers. This paper tackles the challenges associated with accelerating attention through a hardware-software co-design approach while leveraging compute-in-memory (CIM) architecture. In particular, our energy- and area-efficient CIM based accelerator, named HASTILY, aims to accelerate softmax computation, an integral operation in attention, and minimize their high on-chip memory requirements that grows quadratically with input sequence length. Our architecture consists of novel CIM units called unified compute and lookup modules (UCLMs) that integrate both lookup and multiply-accumulate functionality within the same SRAM array, incurring minimal area overhead over standard CIM arrays. Designed in TSMC 65nm, UCLMs can be used to concurrently perform exponential and matrix-vector multiplication operations. Complementing the proposed architecture, HASTILY features a fine-grained pipelining strategy for scheduling both attention and feed-forward layers, to reduce the quadratic dependence on sequence length to linear dependence. 

Further, for fast softmax computation which involves computing the maxima and sum of exponential values, such operations are parallelized across multiple cores using reduce and gather strategy. We evaluate our proposed architecture using a compiler tailored towards attention computation and a standard cycle-level CIM simulator. Our evaluation shows end-to-end throughput ($TOPS$) improvement of $4.4\times$ $- 9.8\times$ and $1.7\times$ $- 5.9\times$ over Nvidia A40 GPU and baseline CIM hardware, respectively, for BERT models with INT-8 precision. Additionally, it shows gains of $16\times$ $-36\times$ in energy-efficiency ($TOPS/W$) over A40 GPU and similar energy-efficiency as baseline CIM hardware. Code base of our evaluation setup will be open-sourced at github.
\end{abstract}


\begin{figure}[ht]

    \centering
    \includegraphics[width=0.9\columnwidth, trim={0 4mm 0 0 0}]{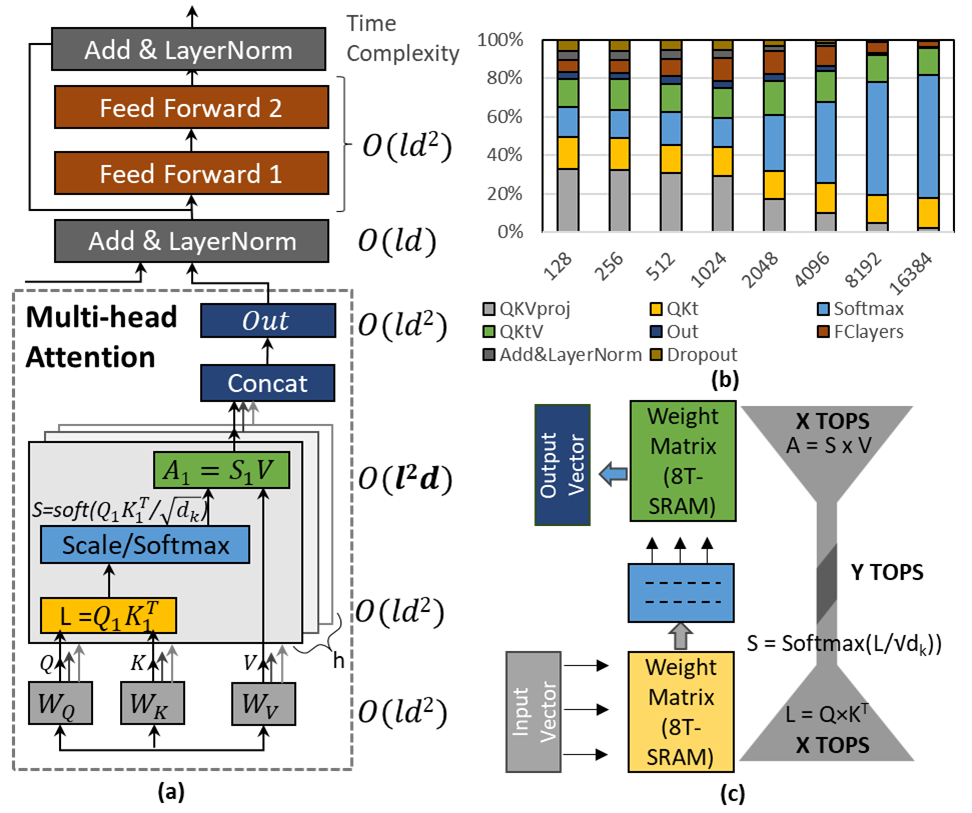}
    \caption{(a) Illustration of a transformer block showing multi-head attention and feed-forward blocks. \cite{AttentionIsAllYouNeed} (b) Runtime breakdown for different sequence lengths and embedding size of 512 as measured on Ampere (A40) GPU \cite{Ampere}. (c) Dependence of softmax on attention computation, limiting the overall throughput of CIM based hardware.}
    \label{fig:intro1}
\end{figure}

\section{Introduction}
\IEEEPARstart{T}{he} advent of transformers has fundamentally transformed deep neural networks (DNNs), enabling significant advancements in natural language processing (NLP) \cite{conneau2019cross, dai2019transformer, AttentionIsAllYouNeed} and computer vision \cite{swinTransformer, dosovitskiy2021imageworth16x16words, bello2020attentionaugmentedconvolutionalnetworks, chen2020generative, wu2021cvt, sun2020transtrack}. Behind the success and widespread adoption of transformers is the \textit{scaled dot-product attention} \cite{AttentionIsAllYouNeed}, which captures relationships and dependencies within long input sequences effectively. As a result, a plethora of recent research has focused on accelerating attention to meet their growing applications demands \cite{A3, Spatten, EdgeBERT, lu2021sanger, softermax, dao2022flashattention, kao2023flat, kwon2023efficient, zadeh2020gobo, softmax1_hyft, softmax2_ITA, softmax3}.

Attention computation consists of memory-intensive matrix-multiplication (MatMul) - softmax - MatMul operation as demonstrated in Fig. \ref{fig:intro1}(a). Specifically, it involves computing \textit{logits} ($QK^T$) from the MatMul between reference \textit{query} ($Q$) and context \textit{keys} ($K$). The logits are then scaled ($\sqrt{d_k}$)  and passed through softmax operation, and finally used in a MatMul with the \textit{value} ($V$) to produce the attention score, \textit{softmax($QK^T/\sqrt{d_k}$)$V$}. 
In the runtime distribution of a transformer block shown in Fig. \ref{fig:intro1}(b), MatMul and softmax operations show comparable execution times for lower sequence length ($l <= 1024$). 
However, as the sequence length ($l$) increases, the softmax runtime grows disproportionately, highlighting the need for its optimization along with MatMul. 
In other words, the presence of softmax operation at the heart of attention makes accelerating transformers distinct from other compute-intensive neural networks such as convolutional and batched fully connected layers. 
Additionally, attention computation requires storing large intermediate matrices ($QK^T$) whose size increases quadratically with the sequence length ($l^2$) \cite{kao2023flat, dai-etal-2019-transformer, choromanski2022rethinkingattentionperformers}. The intermediate matrix size imposes high on-chip memory requirements on the hardware during runtime. Therefore, accelerating transformers needs to solve two main challenges: 1)  the intricate dependency of attention on softmax, and 2) the large memory footprint of intermediate matrix.

\begin{figure}[t!]
    \centering
    \includegraphics[width=0.9\columnwidth]{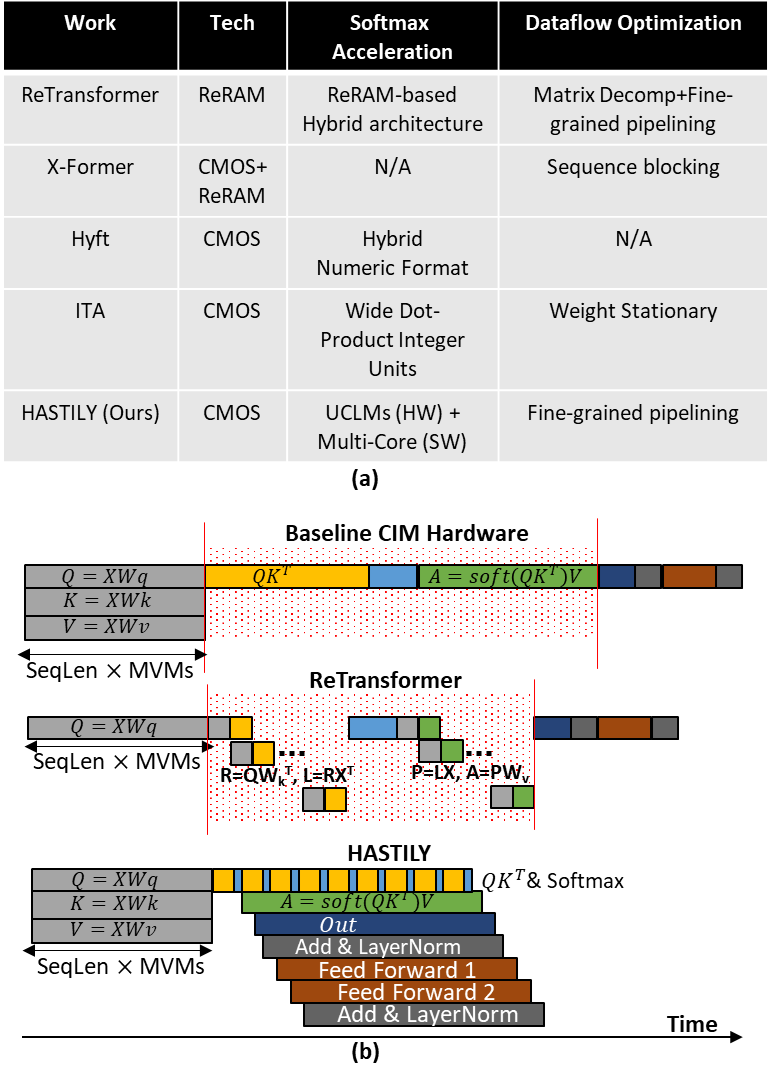}
    \caption{(a) Comparison of our work, HASTILY, with other works \cite{retransformer, xformer, softmax1_hyft, softmax2_ITA} addressing the challenges associated with softmax computation. (b) Distinguishing the fine-grained pipelining technique proposed in this work compared to a prior work, ReTransformer \cite{retransformer}.}
    \label{fig:intro2}
\end{figure}


 Owing to the above-mentioned compute complexities and high on-chip memory requirements of transformer, multiple studies have focused on algorithm-hardware co-design and dataflow optimization techniques. While co-design approaches utilize approximate softmax or pruning to reduce the compute complexity of attention, dataflow optimizations use tiling and tensor-fusion to reduce the on-chip memory footprint \cite{A3, Spatten, EdgeBERT, lu2021sanger, softermax, dao2022flashattention, kao2023flat, kwon2023efficient, zadeh2020gobo}. On the other hand, the limitations in traditional computing technology such as the slower growth rate of memory bandwidth and capacity, known as the ``memory wall", demands for new computing paradigms \cite{wulf1995hitting, gholami2024ai}. In the past, compute-in-memory (CIM) has shown great promise in accelerating convolutional and fully-connected layers by leveraging their weight stationary nature, as their feature maps are known before runtime \cite{verma2019memory, Isaac, aliCIM}. However, attention includes dynamic weights (logit/attention) which leads to an extra data loading cost in terms of latency and energy, limiting the benefits of CIM. Therefore, prior CIM works on transformer acceleration focused on solving such data loading overhead using dataflow optimizations or on accelerating softmax. For instance, ReTransformer \cite{retransformer} proposed a hybrid softmax unit using high-density resistive memory (ReRAM) and decomposes the dot-product matrix into two sub-matrices to eliminate the need to write the whole logit matrix before performing MatMul in CIM hardware. Other CIM works focused on content addressable memories for softmax, ReRAM-SRAM hybrid architecture, and processing in off-chip memory to facilitate efficient data communication during transformer inference \cite{imact, xformer, transpim}. Some research studies have also focused only on designing an efficient softmax unit for faster attention computation \cite{softmax1_hyft, softmax2_ITA, softmax3}, but that comes at an increased area cost. Moreover, the mentioned hardware accelerators face limitations in end-to-end applications due to the need of storing large intermediate matrices for softmax computation.
 
This work fills the gap in CIM based hardware for transformers by proposing a novel hardware-software co-design approach, HASTILY. Our architecture efficiently accelerates the whole transformer block, including softmax and matrix-multiplication, while minimizing area overhead, as shown in the comparison with other works (Fig. \ref{fig:intro2}). In addition, while our software techniques are devised to complement the proposed SRAM-based analog CIM hardware, we believe our compiler can be easily utilized to evaluate other CIM proposals for transformers.
 First, we propose and design an area- and energy-efficient unified compute and lookup module (UCLMs) that constitute the core of our spatial hardware architecture. UCLMs consist of 8T-SRAM arrays for performing both compute and lookup operations within the same memory, whose details are provided in Section \ref{sec:core}.
 The additional functionality of lookup is achieved by just adding an extra source line to the memory array, resulting in no area overhead as observed from the implementation done using TSMC 65nm technology. The dual-functionality of the proposed SRAM array helps in concurrent execution of matrix-vector multiplications and exponential operations across multiple cores. Other than exponent operations, softmax computation also involves finding the maxima and reduction operations to calculate a numerically stable value, as explained in background Section \ref{sec:preliminaries}. To reduce the latency of such calculations and keep hardware changes minimal,
  computations of the local maxima and partial sums of exponential values are parallelized across multiple cores in software and then gathered to a single core in a binary tree fashion.

 Second, we devise a fine-grained pipelining strategy to tackle the need for writing large intermediate matrix in memory during attention. As illustrated in Fig. \ref{fig:intro2}(b), our strategy divides the attention and feed-forward computation into fine-grained pipeline where one vector of input matrix is processed in a single stage of the pipeline. The underlying concepts behind the success of this approach are the concurrent execution of logit calculation and softmax in hardware and maintaining all data dependencies between different MatMul operations in software. 
 By keeping the pipeline fed by a vector instead of a matrix during runtime, as explained in Section \ref{sec:compiler}, we are able to reduce the space complexity dependence on sequence length from quadratic to linear.
 
In summary, the contributions of our work can be listed as follows:
\begin{itemize}
    \item We propose a novel softmax-friendly high-throughput compute-in-memory architecture with minimal area overhead for accelerating attention. It consists of 8T-SRAM arrays (TSMC65nm) acting as unified compute and lookup modules (UCLMs). 
    \item The proposed architecture solves the softmax bottleneck by enabling concurrent matrix-vector multiplication and exponential operations in hardware. Further, the maxima and reduction operations are accelerated in software to keep the time complexity of the order of $\mathcal{O}(log n)$.
    \item We further reduce the on-chip memory requirement for attention calculation by incorporating a finer-granularity pipelining strategy, processing transformer computation in smaller blocks while maintaining all data dependencies. 
    \item The evaluations, done using a compiler tailored for transformer computations and a standard CIM cycle-accurate simulator, demonstrate throughput improvements of up to 9.8$\times$ over Ampere GPU and 5.9$\times$ over the baseline CIM hardware for BERT models.  
\end{itemize}



\begin{table*}[t!]
    \centering
    \caption{Comparison between different ways of implementing exponential function}
    \label{tab:exponent}
    \begin{tabular}{|c|c|c|c|c|} \hline 
         \textbf{Characteristics}&  \textbf{Software}& \textbf{ LUT on-chip}&  \textbf{LUT loaded from off-chip}& \textbf{HASTILY}\\ \hline 
         Area overhead&  No&  Yes&  No& No\\ \hline 
         Power consumption&  Normal&  Low&  High& Low\\ \hline 
         Compute Latency&  High&  Low&  High (only loading)& Low\\ \hline
    \end{tabular}

\end{table*}

The details of the evaluation methodology are given in Section \ref{sec:methodology}, followed by results and discussion in Section \ref{sec:results} and Section \ref{sec:discussion} respectively. The final Section \ref{sec:conclusion} concludes the paper.

\section{Preliminaries and Challenges}
\label{sec:preliminaries}

This section briefly discusses the background concepts related to transformer models and CIM based hardware. We also highlight the unique challenges associated with accelerating transformers and current limitations of CIM based hardware accelerators.

\subsection{Transformer models}

A transformer network comprises multiple blocks and a final classifier, where each block can further constitute of an encoder block, decoder block, or a combination thereof. Encoder-like architectures, such as the bidirectional encoder representation (BERT) models, process the entire input sequence in one pass \cite{devlin2018bert}. As the name suggests, BERT utilizes a bi-directional attention mechanism where each token can attend to all other tokens in the sequence. Such models are used in applications like classification \cite{sun2020finetuneberttextclassification, bertTextClassification} and search engines \cite{kocián2021siamesebertbasedmodelweb}. Conversely, decoder-like architectures, such as the generative pretrained transformer (GPT) models, process input sequentially, with each token attending only to previous tokens through triangular masking during attention calculation \cite{yenduri2023gpt}. Tasks such as summarization \cite{goyal2023newssummarizationevaluationera, automatedNewsSummarization} and text prediction \cite{thoppilan2022lamdalanguagemodelsdialog, gpt-j, zhang2022optopenpretrainedtransformer} are common applications for GPT models. Recently, encoder-like transformer models have also been adopted for vision tasks, as demonstrated by the latest vision transformer models \cite{you2023vitcod, zhu2023vlgptgenerativepretrainedtransformer}. The key operations in transformers can be categorized as matrix-multiplication (or dot-product), softmax and other element-wise operations, whose details are mentioned below.

\subsubsection{Matrix Matrix Multiplication Operations}
Matrix multiplication (\verb|matmul|) operations in a transformer layer can be categorized into dynamic and static types, based on whether the weight matrix is calculated during runtime or known beforehand. For that reason, dynamic MatMul is also commonly referred as activation - activation MatMul operation. Static MatMuls include the $Q/K/V$ projection, $Out$ projection, and $Feedforward$ layers, with a compute complexity of $O(ld^2)$ as shown in Fig. \ref{fig:intro1}(a). Dynamic MatMuls involve the $QK^T$ and $SV$ computations with a time complexity of $O(l^2d)$. Here, $l$ denotes input sequence length and $d$ is the size of embedding dimension. Since matrices are generated during runtime in dynamic MatMuls, and CIM hardware benefits heavily from its weight stationary nature, such MatMuls limit the benefits of CIM unless properly addressed. Also, the \textit{quadratically increasing size of intermediate matrix} increases the on-chip memory capacity demands and data movement costs during transformer inference.

\subsubsection{Softmax Function}
Softmax exponentiates the values of scaled logit matrix $L=QK^T/\sqrt{d_k}$ for each input sequence vector and facilitates in calculation of final attention score. Attention score decides the probability of predicting the next token in the sequence. Generally, softmax is calculated as follows to avoid overflow conditions:
\begin{equation}
    \text{softmax}(M_{i,j}) = \frac{\exp(M_{i,j}-max_j)}{\sum_{j} \exp(M_{i,j}-max_j)}
    \label{eq:softmax}
\end{equation}
Therefore, calculating the softmax includes multiple element-wise or vectorized operations:
\begin{itemize}
    \item Maxima: Calculating the maxima among all elements in a vector
    \item Subtraction: Elementwise subtraction of calculated maximum from the vector
    \item Exponent: Finding exponent values for each element in the vector
    \item Reduction: Sum of all exponents within a vector to calculate denominator
    \item Division: Elementwise division (or reciprocal multiplication) by the sum
\end{itemize} 

Other than the exponent, the element-wise operations are simple to perform using standard arithmetic and logic units (ALUs). The exponential function can be calculated as an infinite sum using the MacLaurin series expansion: 
\begin{equation}
    e^x=1+x+ x^2/2!+x^3/3!+\cdots + x^n/n! + \cdots.
\end{equation}
Since calculating the sum of infinite terms is cumbersome, an approximate exponential value is calculated by dropping the higher order terms and taking only the sum of the first few terms. 
On one hand, a fully software implementation of softmax offers a low-cost and high-latency solution where the exponential values are calculated during runtime. On the other hand, for faster exponential computation, look-up tables (LUTs) can be stored on-chip to directly access the pre-calculated values, costing extra area. It is also possible to load the LUTs on chip during runtime for limited area constraint hardware as summarized in Table \ref{tab:exponent}.
However, \textit{instantiating large lookup tables leads to large power overheads}, as also mentioned by \cite{softermax}. 

\subsubsection{Other Element-wise Operations and Functions}
Each transformer block adds the output from the previous block to the multi-head attention and feedforward layer output. Additionally, non-linear functions such as GELU (Gaussian Error Linear Unit) and LayerNorm (Layer Normalization) are employed in transformer models. GELU provides a smoother activation function compared to ReLU, maintaining a non-zero gradient at $x=0$. LayerNorm normalizes inputs using their mean and variance before adding the previous layer's output. Although such functions are essential, the runtime of GELU and LayerNorm is not as significant compared to matrix multiplications and softmax operations because they operate on comparatively smaller tensors, as shown in Fig. \ref{fig:intro1}(b).

\subsection{Compute in memory (CIM) based Hardware}\label{sec:cim_hardware}

Compute in memory (CIM) enables matrix-vector multiplication operation inside a memory array by modifying the peripherals. The computational nature of CIM can be analog or digital depending on the associated peripherals and how multiply-and-accumulation (MAC) is performed. Analog CIM typically activates multiple wordlines in a memory array and accumulates current from all rows in the bit line (BL) to produce analog MAC output. The activation of wordlines is decided based on the input vector by incorporating Digital-to-Analog Converters (DACs) in the memory row decoders. Additionally, the BL analog output is converted to digital using Analog-to-Digital Converters (ADCs) and then shifted-and-added (S\&A) with partial sums from other columns to compute the final matrix-vector multiplication output \cite{analog1cim, twin8t}. Digital CIM typically employs bit-wise arithmetic logic such as a NAND gate to perform bit-wise multiplication of inputs, and then combines the partial results to compute the final output \cite{computesram, tmsc2023digital}.

In terms of memory technology, analog CIM can be implemented using either SRAM (static random access memory) or emerging memories such as resistive RAMs (ReRAMs) or magnetoresistive RAMs (MRAMs). While emerging memories offer a high-density memory solution, they typically suffer from issues such as resistance drift, write endurance and/or low on-off distinguishability \cite{chakraborty2020resistive}. In this work, we focus on SRAM based CIM hardware because of their higher feasibility in terms of implementation.

Analog CIM hardware offers a highly energy-efficient solution for accelerating neural networks due to their efficient execution of matrix-vector multiplication in analog form. The initial works on analog CIM based accelerators focused on convolutional and fully connected networks \cite{prime, Isaac, puma, song2017pipelayer}. Such spatial accelerator architectures often feature a weight-stationary dataflow and low-precision computations such as 8-bit or 16-bit datawidth to achieve higher energy efficiency. However, when analog CIM is used for transformers, the presence of dynamic MatMuls reduces their energy-efficiency because activation-activation MatMuls require writing the matrix into memory before performing \verb|matmul|, thereby reducing their weight stationary benefits. Note that SRAM based CIM will have reduced data loading overhead during dynamic MatMul operations due to their lower write energy/latency compared to emerging memory technologies \cite{agrawal2021magnetoresistive}.

\begin{figure*}[ht!]
    \centering
    \includegraphics[width=\textwidth, trim={0 6mm 0 0 0}]{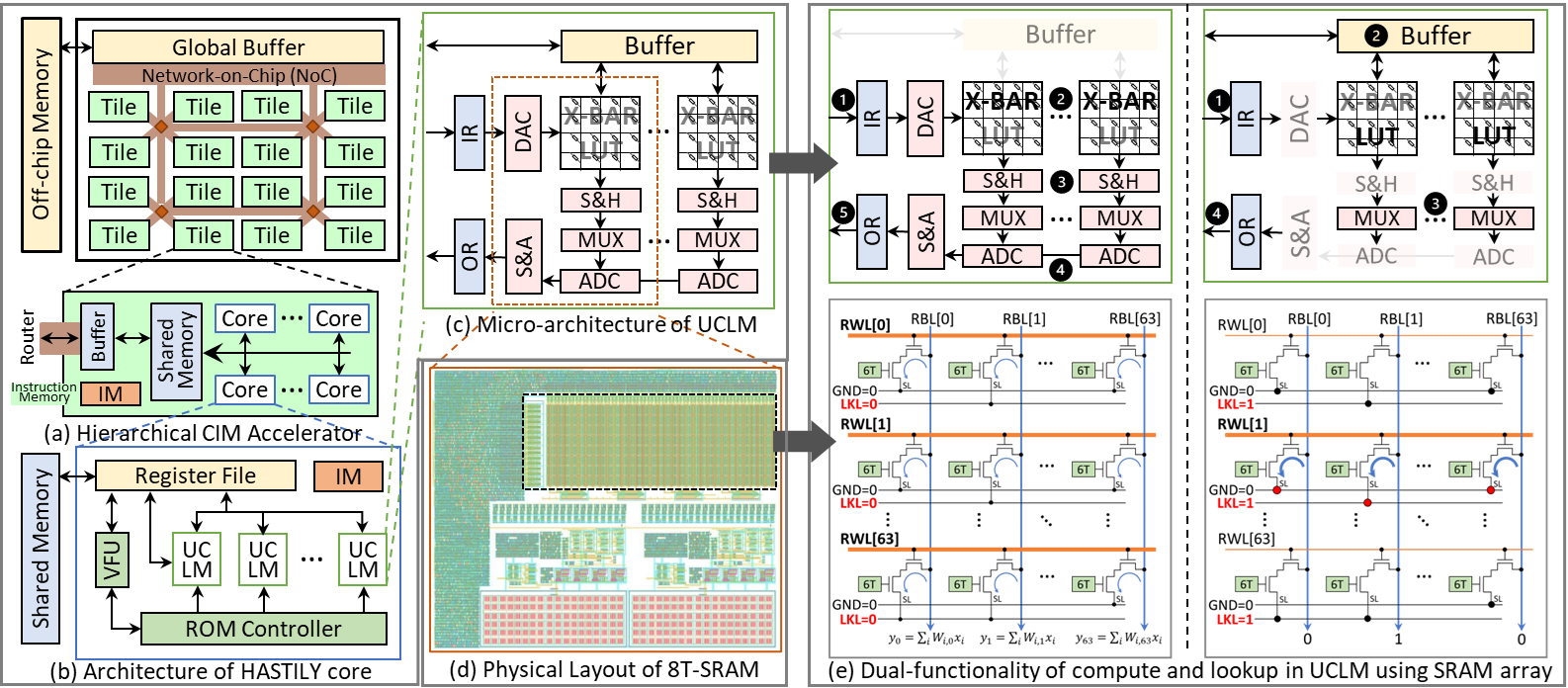}
    \caption{(a) The hierarchical spatial architecture of the proposed CIM accelerator, (b) hardware architecture of each core in HASTILY, (c) micro-architecture details of each UCLM consisting of multiple SRAM arrays, (d) physical layout of each dual-functionality 8T-SRAM array implemented in TSMC65nm and (e) depiction of the two operations, compute on the left and lookup on the right, in UCLM and correspondingly in the SRAM.}
    \label{fig:main_arch_cim}

\end{figure*}

To maximize the throughput, CIM based spatial accelerators constitute of a hierarchical architecture as proposed in a previous work, PUMA \cite{puma}. At the chip level, the architecture consists of multiple tiles and a Global Buffer (GB), as illustrated in Fig. \ref{fig:main_arch_cim}(a).
GB serves as a cache bridging off-chip memory and the tiles, while tiles communicate with each other via send/receive instructions.
Each tile consists of multiple cores, shared memory, and an instruction memory. 
Cores perform computations and access data stored in shared memory. \textit{The size of shared memory could limit the throughput benefits of CIM cores if it cannot accommodate the intermediate matrix}, resulting in off-chip accesses and increasing compute latency.

Typical CIM core constitutes of multiple Matrix-Vector Multiplication Units (MVMUs), a register file, a Vector Function Unit (VFU), and instruction memory. The core executes instructions such as loading and storing data from/to shared memory and performing matrix-vector multiplications (MVMs) within the MVMUs. MVMs operate in SIMT (Single Instruction Multiple Thread) mode, where multiple MVMUs generate output vectors concurrently. The VFU handles arithmetic operations, including element-wise addition, multiplication, division, non-linear functions (like ReLU), normalization, and transcendental functions (like exponential). However, \textit{VFUs cannot efficiently handle the high frequency and complexity of softmax operations in transformers}, limiting the overall throughput of CIM based accelerators for transformer models. Additionally, incorporating a large VFU could potentially reduce the softmax computation time, but adds significant area overhead.

\section{Proposed CIM Core Architecture}
\label{sec:core}

The hardware architecture of HASTILY constitutes of a spatial accelerator consisting of an array of tiles, each containing multiple cores and shared memory as shown in Fig. \ref{fig:main_arch_cim}(a). 
Each core features multiple unified compute and lookup modules (UCLMs) for high-throughput execution of matrix-multiplication (\verb|matmul|) and exponent (\verb|exp|) operations. Fig. \ref{fig:main_arch_cim}(b) illustrates that each core also contains a vector functional unit (VFU) for vectorized arithmetic operations, register file (RF) for feeding data to the UCLMs and a lookup (LK) controller. The micro-architecture of UCLM shown in Fig. \ref{fig:main_arch_cim}(c) consists of dual-functionality 8T-SRAM CIM arrays, that can perform MAC operations and memory lookup operations. LK controller helps in switching the dual-functionality SRAM array mode between compute and lookup.

\subsection{Dual-functionality 8T-SRAM arrays}
 Other than compute, the lookup functionality in 8T-SRAM array is achieved with the addition of a lookup line (LKL) at each row in the array. While the cross-coupled inverter in SRAM cell decides the weight matrix data, the lookup-table data is stored based on the connection of sourceline (SL) with LKL or ground (GND)  \cite{fong2015embedding, lee2012area}. 
 The layout of such dual-functionality array implemented in TSMC 65nm technology is shown in Fig. \ref{fig:main_arch_cim}(d), occupying same area as standard CIM array. Although the extra source line (LKL) does not add any extra area to the array, the peripheral circuits are slightly modified to enable lookup operations along with the crossbar operation. The details of how the different operations are executed in the SRAM array is explained below.

\subsubsection{Read/Write operation} For read operation, first the read bitlines (RBLs) are pre-charged, followed by activating the read wordline (RWL) to select the corresponding row. Then, RBL gets discharged depending on the data stored in the SRAM cell and the final voltage is read at the column peripheral. Similarly, for write operation, corresponding write wordline (WWL) is activated and write bitline (WBL/WBLB) is driven high or low depending on the data that needs to be written to the memory cell. Note, that LKL remains grounded during SRAM read/write operations. 

\subsubsection{Lookup operation} To utilize the SRAM array for exponential table lookup operation, first LKL is set to zero and the corresponding row of SRAM is copied to a temporary buffer by asserting the RWL. Next, the write wordline (WWL) is asserted to write '1' to the SRAM row. After these initial 2 cycles, LKL is set to high and read wordline (RWL) is asserted to read the data from the row through the read bitlines (RBL). 
If SRAM cell’s SL is connected to LKL, RBL outputs '1'; otherwise, if it is connected to GND, RBL gets discharged and outputs '0'.
Since the buffer stores the data originally present in the SRAM row, WWL is asserted again to write the data from the buffer back to the row to preserve the original information. While the data is being written back from the buffer, multiplexer sends the required lookup table value to the output register. In this manner, each SRAM array can be used for lookup operations with no area overhead and with a latency of 4 cycles. 

\subsubsection{Compute operation} During SRAM compute operation, LKL is kept low so that all SLs are grounded. The compute operation begins with pre-charging BL to the supply voltage (VDD) using an active low pulse. Multiple RWLs are activated (set to high) simultaneously, and the BL discharges proportional to the dot-product between the input vector (represented by the voltages on the word lines) and the corresponding weight column vector (stored in the memory cells). The resulting voltage on the BL, now reflecting the computed dot-product, is sampled by the sample-and-hold (S\&H) block. Analog multiplexers are used to select which columns are active for computation at any given time. The ADC converts the analog voltage into a digital signal, which is further processed by the shift and add unit (S\&A). S\&A unit manages different bit significance for inputs and weights, computing the final MAC output.

\begin{figure*}[ht]
    \centering
    \includegraphics[width=0.85\textwidth, trim={0 7mm 0 0 0}]{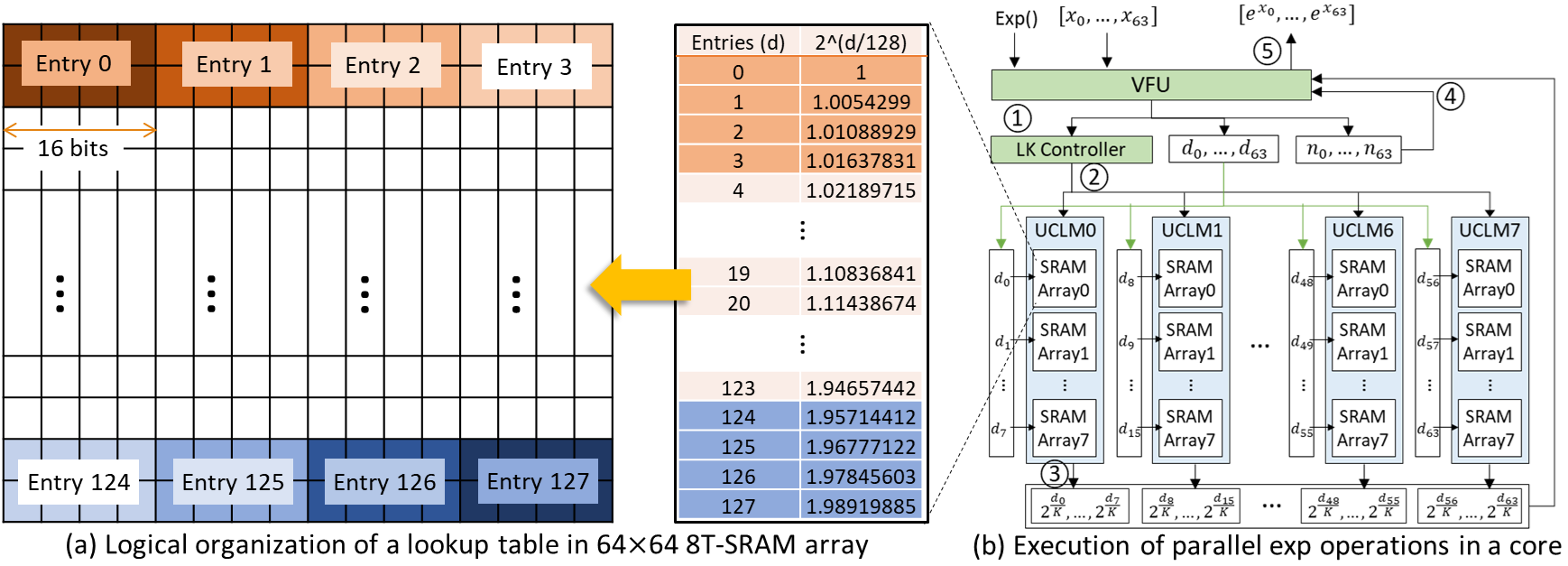}
    \caption{(a) Each SRAM array stores a 128-entry lookup table for $2^k/128, 1\leq k\leq128$ (b) Illustration of parallel exponent operations execution in multiple UCLMs in a core of HASTILY architecture. }
    \label{fig:element-wise-exp}
\end{figure*}

\begin{figure}[ht]
    \centering
    \includegraphics[width=\columnwidth, trim={0 9mm 0 0 0}]{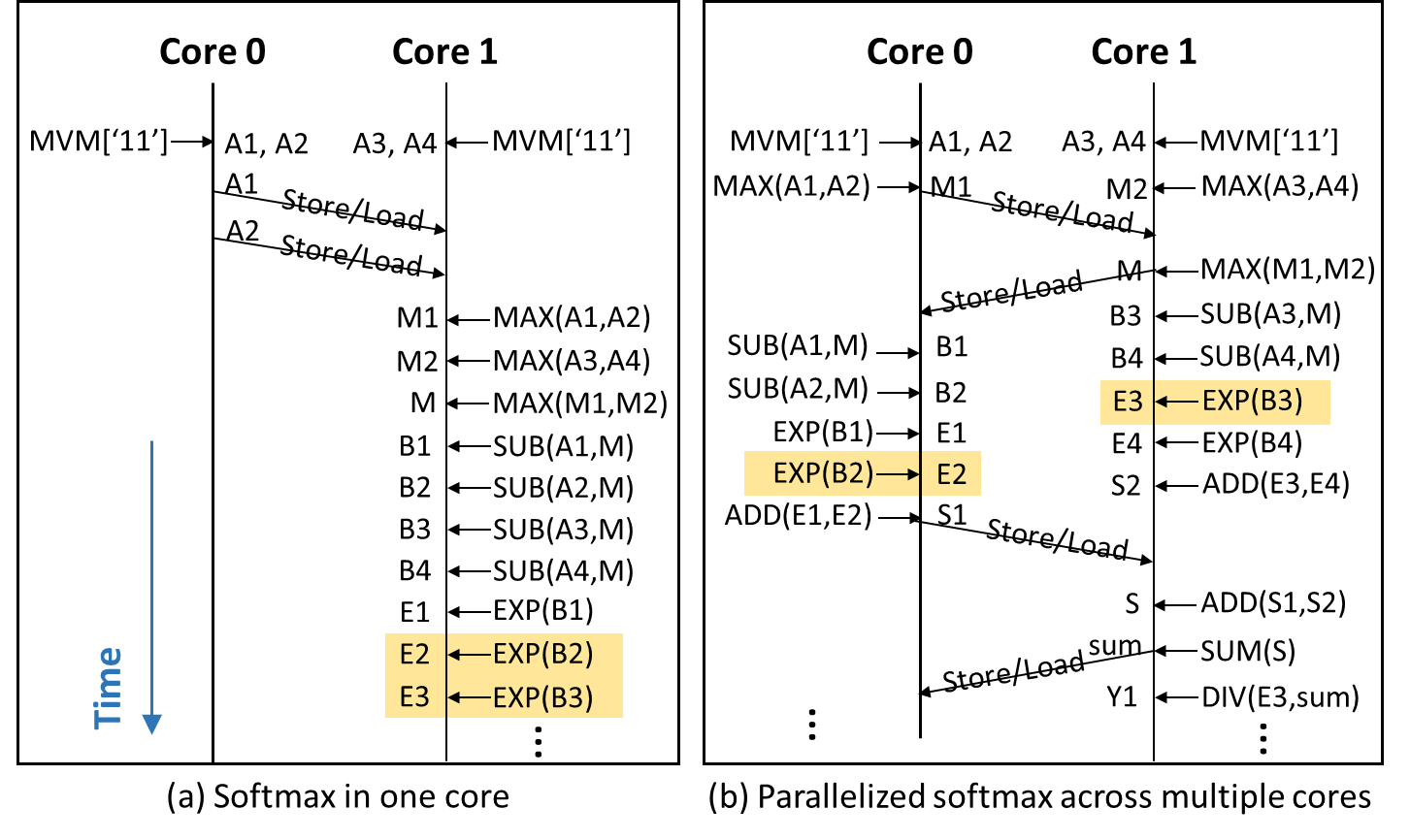}
    \caption{Computing softmax by gathering all required vectors in a single core versus parallelizing the compute across multiple cores and gathering them in a tree-like fashion.}
    \label{fig:parallel_max}
\end{figure}
\subsection{Hardware-Software Co-design for Softmax}

As described in Section \ref{sec:preliminaries}, softmax computation involves various vectorized and/or element-wise operations. We utilize the capability of dual-functionality SRAM arrays to store lookup tables for exponent operation (\verb|exp|) and optimize other operations such as \verb|max| and \verb|add| in software. Here, we explain the details of such optimizations and the proposed micro-architecture for high-throughput execution of softmax operation.

\subsubsection{\textbf{UCLMs} for parallel exponent operations} Softmax requires finding the exponent of each element in a vector among other operations, as depicted in Equation \ref{eq:softmax}. To simplify the computation of exponential function, it can be transformed into the following form \cite{intelHarrison1999TheCO}:
\begin{equation}
    e^x = 2^n \times 2^{d/K} \times e^r  
\label{eq:exp-decomp}
\end{equation}
where $K$ is a chosen integer (representing the number of entries in a pre-computed lookup table), $n=\lfloor x/log(2) \rfloor$, $d=\lfloor (x/log(2)-n)K \rfloor$, and $r=x-n log(2)/K$. 
Therefore, the exponent value can be simply calculated with a combination of a table lookup, bit-shift operations and multiplication by a constant. Such a lookup table will contain values from $2^0$ to $2^{(K-1)/K}$ to directly fetch the value of $2^{d/K}$ and $n$ will decide the bit-shift operation. The residual term $e^r$ is approximated with MacLaurin series expansion, but due to its small range $(1 \leq e^r < 1+1/K)$, it can be approximated as 1 or 1+r with minimal error. In this work, we store a pre-computed table for storing the $2^{i/K}$ entries with $K=128$ in the SRAM array. 
The approximation of $K=128$ yields an error rate below 0.54\% and 0.0015\% with 1 and 1+r, respectively.

Considering a bit-precision of 16-bit for each value in the 128-entry exponent table, we propose to organize the exponent table in our dual-functionality SRAM array of size $64\times64$ as illustrated in Fig. \ref{fig:element-wise-exp}(a). Since each UCLM contains multiple SRAM arrays, each core is capable of executing multiple exponent operations in parallel. Fig. \ref{fig:element-wise-exp}(b) illustrates how an \verb|exp| operation for a vector containing 64 elements is executed in parallel in the HASTILY core. 
To execute a vectorized \verb|exp| operation, the VFU first computes $n$ and $d$ in parallel, then element-wise table-lookups are performed simultaneously using multiple SRAM arrays. Finally, the VFU applies element-wise bit-shifting to complete the operation.

\subsubsection{Optimizing \textbf{Multi-Core} execution}
Moreover, HASTILY can handle longer sequence lengths by parallelizing the exponent operation over a vector to multiple cores. 
Despite parallel execution of operations such as \verb|exp|, \verb|sub|, and \verb|div| in each core, some operations such as \verb|max| and \verb|reduce| require combining partial results from each core while maintaining consistency. Such sequential gathering of partial maxima or sum could increase the total computation time due to contention in the VFU of a single core.

To fully parallelize the softmax operation, we utilize the private VFU in each core to calculate the partial sums and local maxima. Then, we gather and operate (reduce or calculate maxima) the partial values from different cores in a tree-like hierarchy. The multi-core execution of softmax as depicted in Fig. \ref{fig:parallel_max} reduces the time complexity of \verb|max| and \verb|reduce| from $N$ to $log(N)$. While some delay is caused by extra load and store operations, it is minimal compared to fully serialized VFU computation in a single core. Additionally, it becomes a suitable choice to compute softmax across multiple cores when the size of $QK^T$ becomes large, and the $K^T$ matrix already requires multiple UCLMs in a single core to be fully mapped.

An exemplary scenario of the operations happening in two cores during softmax computation is provided in Fig. \ref{fig:parallel_max}(b). For instance, core 0 produces partial vectors $A1$ and $A2$ from MVM computations, while core 1 produces partial vectors $A3$ and $A4$. Each core independently computes the maximum value of its respective vectors, yielding $M1$ and $M2$. These maxima are then gathered to determine the overall maximum value ($M$). The overall maximum is distributed back to the cores through store/load operations in shared memory, allowing each core to perform subtraction and \verb|exp| computation simultaneously. Reduction operation is performed in a similar fashion across multiple cores to compute the denominator for the softmax output.

\section{Fine-grained Pipelining}
\label{sec:compiler}

Pipelining is a standard technique that divides the overall computation into smaller stages, allowing multiple stages to process different data in parallel. In the context of transformers, computation is typically divided into blocks of matrix-multiplication, softmax, and other operations, as illustrated in Fig. \ref{fig:intro1}(a). Our focus is to further divide each of these computations into finer-grained vectors, enabling parallel processing of different vectors over various operations. 

\subsection{Pipelining Matrix-Multiplications}
The key idea is to divide each computational stage into the number of rows of the input matrix, allowing each stage to run in parallel with other stages on a spatial architecture containing multiple tiles and cores. 
Matmul can be executed as iterative MVMs, each MVM computes the dot product between a row of the input and the weight matrix, producing a corresponding row of the output. 
For consecutive MatMuls, the output row from one MVM in the first MatMul can be directly used as input for an MVM in the second MatMul. This allows the subsequent MVM in the first MatMul to run concurrently with the second MatMul, enabling fine-grained parallelism. 
This approach enhances both speed and memory efficiency: speed is nearly doubled due to parallelized computation, while memory requirements are reduced, as storing the entire intermediate matrix for the second MatMul is unnecessary, with only temporary storage for the current vector being needed.

\begin{figure}[!t]
    \centering
    \includegraphics[width=\columnwidth, trim={0 4mm 0 0 0}]{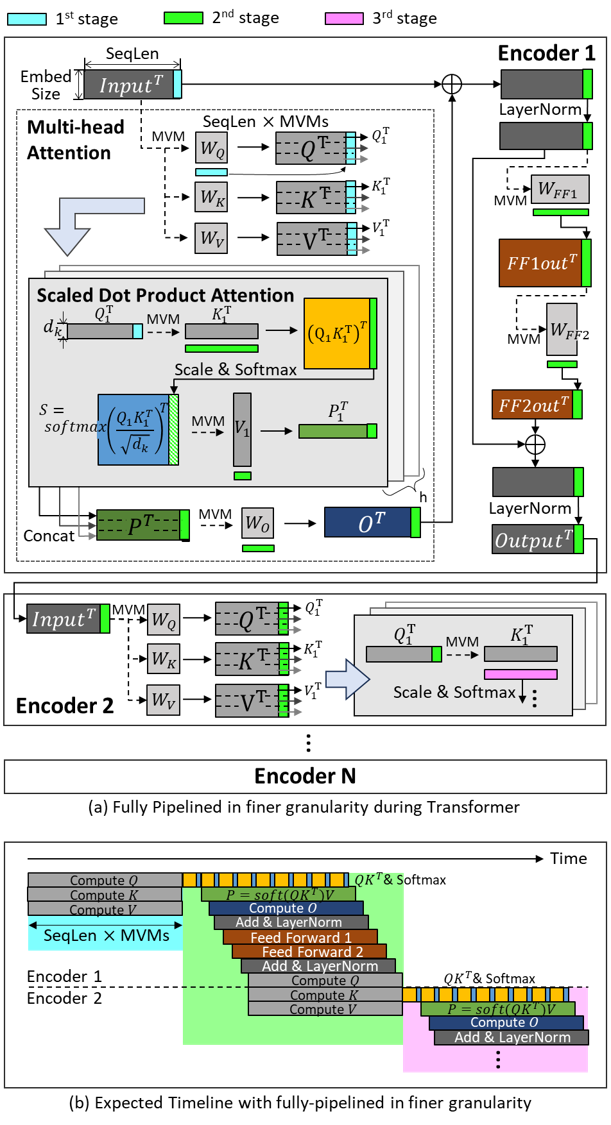}
    \caption{(a) Illustration of the dataflow for scheduling encoder layers of a transformer model using the fine-granularity pipelining technique and (b) the corresponding timeline of different stages shown in unique colors.}
    \label{fig:fine-gran-timeline}
\end{figure}
\subsection{Pipelining Transformer}

Fig. \ref{fig:fine-gran-timeline}(a) illustrates the pipelined flow of matrices and intermediate vectors  during transformer inference and (b) shows the pipelined timeline. Note that in the figure, inputs and outputs of MVMs are depicted as transposed to represent the inputs entering the word lines of the SRAM array during MVMs. However, no actual transpose operation is performed on the inputs or outputs, except for $K^T$, which is transposed and used as a weight matrix.  

\subsubsection{Q,K,V Projection}
The input is simultaneously multiplied by the three matrices, $W_Q$, $W_K$, and $W_V$, with each matrix performing $seqLen \times$ MVMs. Once the $Q$, $K$ and $V$ matrices are generated, $K$ and $V$ are written to the next cores' UCLMs as weights, labeled $K_i$ and $V_i$, where $i$ corresponds to the number of attention heads. 
Before being written to the UCLM arrays, $K_i$ is transposed ($K_i^T$) within the allocated cores by modifying the matrix's address mapping. 
Each head then performs scaled dot-product attention in parallel, starting with $Q_iK_i^T$. However, the $Q_i/K_i$ projection block cannot be pipelined with the $Q_iK_i^T$ operation because the complete $K_i$ matrix must be available before the computation of $Q_iK_i^T$. This marks the end of the first stage of the process (highlighted in blue in Fig. \ref{fig:fine-gran-timeline}).

\subsubsection{Pipelining Attention}
The second stage (highlighted in green) operates in a fully-pipelined manner. Each column of the generated $Q_i^T$ is fed sequentially into the mapped $K_i^T$ within the UCLMs, generating the head output $(Q_iK_i^T)$, one vector at a time. This output vector is then scaled by dividing by $\sqrt{d_k}$ and processed through softmax operations within the same UCLMs using the LT mode. Once the softmax is complete, the resulting vector is immediately passed to the next UCLM to perform multiplication with $V_i$, producing $P_i$. While $P_i=SV_i$ is being computed, the next $Q_iK_i^T$ computation and its associated softmax can proceed in parallel, enabling pipelining. Since all attention heads generate their respective $P_i$ outputs concurrently, each output vector from $P$ can  undergo the linear transformation as soon as it is available, producing the final attention head output $O$ in a pipelined fashion.

\subsubsection{Pipelining within Encoder Layer}
Next, addition of a vector from $O$ with a vector from the input can be pipelined, and the subsequent layer normalization can also be performed immediately since it operates on the result of the addition. After the Add \& LayeNorm, the two feed-forward networks, $FF1Out$ and $FF2Out$, can also be pipelined with finer granularity. Following this, another Add \& LayerNorm is performed to generate the final output of the encoder layer, again in a pipelined manner. 

\subsubsection{Pipelining between Encoder Layers}
The output generated from Encoder 1 serves as the input for Encoder 2. The first output vector of Encoder 1 can be immediately used as the first input vector of Encoder 2 to compute $Q$, $K$, and $V$. Then, the same process as in Encoder 1 is repeated. This marks the end of the second stage, and the third stage (highlighted in purple) begins. To summarize, everything except for the $Q/K/V$ projection can be pipelined to finer granularity. As a result, an encoder consisting of N layers can perform the entire inference calculation in $(N+1) * seqLen \times MVM~time$.


\begin{table}[t]
  \centering
  \scriptsize
  \caption{Hardware Characteristics}
  \label{table:sim-arch}
\setlength{\tabcolsep}{5pt}
\renewcommand{\arraystretch}{1.1}
  \begin{tabular}{|c|c|c|c|c|}
    \hline
    \textbf{Component}&  \textbf{Power(mW)}& \textbf{Area(mm$^2$)} &\textbf{Parameter}&\textbf{Specification} \\
    \hline
    \hline
        Global Buffer&  25.35&  0.242&Capacity&256KB\\
    \hline
    Tile& 1.14K$^*$  &  2.57&\# per node&128\\\hline
    Shared Mem&  25.35&  0.242&Capacity&256 KB\\
    \hline
    Memory Bus&  6&  0.18&width&512 bit \\ \hline 
    Core& 140.3 $^{\dag}$& 0.2755& \# per Tile&8 per Tile\\
    \hline
    VFU&  1.7&  0.004&ALU width&64 \\
    \hline
    Register File&  1.14&  0.0078&Capacity&4 KB \\
    \hline 
    UCLM&  MM$^{\ddagger}$: 22.38&  0.0155&\# per Core    &16 \\
        & LT: 0.518&         & \# of CIMArray & 8 \\ 
    \hline
    DAC&  0.0035&  1.67e-7&\# per CIMArray&64 \\
           & &   & Bit-precision& 1 bit\\
    \hline
    ADC&  1.35&  0.0011&\# per CIMArray& 1 \\
           & &   & Bit-precision& 6 bit\\
    \hline
  \end{tabular}
    \begin{tablenotes}
    \item[1] $^*$Scaled from Core total power and tile memory related instructions.
    \item[2] $^{\dagger}$Includes all UCLMs in a core performing MVM, load/store to/from Register File.
    \item[3] $^{\ddagger}$Includes MVM operation, S\&H, S\&A, ADC and writing to output buffer.
  \end{tablenotes}
  
\end{table}

\begin{table}[t!]
  \centering
  \scriptsize
  \caption{Software Benchmarks}
  \label{table:bert}
  \begin{tabular}{|c|c|c|c|c|}
    \hline
    \textbf{Model} & \textbf{Embedding Size} & \textbf{\#Heads} & \textbf{\#Layers}  & \textbf{Hidden Dimension}\\
    \hline
    BERT-Base& 768& 12& 12& 3072\\
    BERT-Large& 1024& 16& 24& 4096\\
    \hline
  \end{tabular}
\end{table}

\section{Evaluation Methodology}
\label{sec:methodology}
\subsection{Simulator}
We used a cycle-level simulator based on PUMASim \cite{puma}, having a hierarchy of chip, tile, core, and MVMUs in 32nm technology. 
Since PUMA is an RRAM-based CIM architecture, we replaced its MVMUs with our UCLMs. 
In addition, we have considered DRAM and a global buffer associated with the chip for fetching and storing the transformer inputs/weights. Since transformer models produce comparable results with lower precision fixed-point representation, we consider 8 bit precision of inputs and weights for all our evaluations \cite{kim2021bert,dettmers2022llm}. The lookup table resolution is set to 16 bit as described in Section \ref{sec:core}. The energy and area characteristics for the lookup table operation are calculated from an implementation done in TSMC65nm (Fig. \ref{fig:main_arch_cim}(b), and scaled down to 32nm for our simulator. The architecture contains $16$ UCLMs per core, each having $8$ crossbars of size $64 \times 64$ for storing different bits of weights. Other architecture specifications along with the energy and area considerations are described in Table \ref{table:sim-arch}.

\subsection{Compiler}
We develop a compiler supporting BERT inference on a CIM based spatial accelerator. The new compiler supports traditional CIM operations present in a previous CIM accelerator \cite{puma}. In addition, the instruction set architecture (ISA) is extended to support operations such as transpose vector, finding maxima of a vector mapped to multiple cores, tree-sum/reduction, and single instruction multiple exponent.

\subsection{Workloads \& Baseline}
We choose two variants of transformer, BERT-Base and BERT-Large for end-to-end evaluation of throughput and energy-efficiency with the Nvidia Ampere (A40) GPU. The specifications of BERT models are shown in Table \ref{table:bert}.  Further, \verb|bitsandbytes| and \verb|transformers| library from hugging-face are used for GPU evaluations of the BERT models keeping 8-bit precision. The power consumption of GPU is measured using \verb|nvidia-smi| functionality and the idle power is subtracted from the observed power numbers to only compare the dynamic energy. For individual layer evaluations, latency is calculated by implementing multi-head attention in Pytorch and measuring runtime using the \verb|time()| library. 

We also compare the performance and energy consumption of our proposed core with PUMA \cite{puma} core across different sequence lengths and embedding sizes. To ensure fair comparison, we replaced ReRAM-based crossbar arrays in the PUMA core with 8T-SRAM-based crossbar arrays. Otherwise, the hardware architecture of PUMA shares the same configuration as HASTILY except the use of UCLMs in our accelerator. In terms of software, PUMA does not support parallelized \verb|exp| operations and relies only on VFU for softmax computation. 
For all the accelerator evaluations, we assume that the data can be stored on-chip, accessing DRAM only for loading the initial inputs and storing the layer/model output. Under the given hardware configuration, baseline CIM accelerator, PUMA, cannot accommodate a model with sequence length $l$ $> 1024$ without spilling the intermediate matrix to DRAM. Therefore, we evaluate PUMA standalone with $l<1024$ or consider fine-grained pipelining when $l > 1024$ to keep comparisons fair.

\section{Results}
\label{sec:results}

This section compares HASTILY with the Nvidia A40 (Ampere) GPU and PUMA, as baseline CIM accelerator in terms of compute latency and energy consumption. The Nvidia A40, fabricated using Samsung's 8nm process, features a die size of $628.4mm^2$ \cite{Ampere}. In contrast, our UCLM-based accelerator occupies same area as PUMA, which is approximately $330mm^2$ considering 32nm process technology. We evaluate the benefits of HASTILY at different levels -- including only softmax, single encoder layer and end-to-end model performance for different batch sizes.

\begin{figure}[t!]
    \centering
    \includegraphics[width=\columnwidth, trim={0 6mm 0 0 0}]{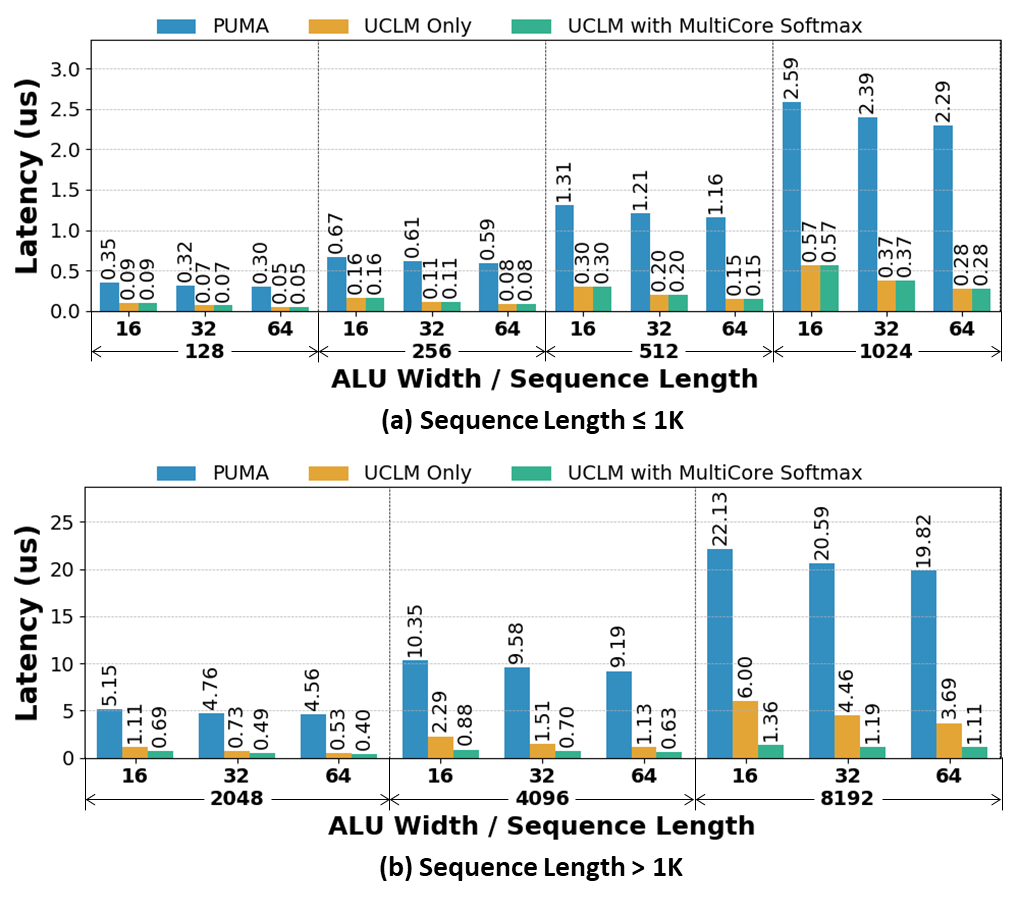}
    \caption{Speed benefits of the proposed UCLM-based hardware and multi-core softmax computation support compared to PUMA. ALU Width refers to the number of ALUs in the VFU.}
    \label{fig:softmax-results-latency}
\end{figure}

\subsection{Softmax Acceleration}

This sub-section presents the performance of our proposed UCLM in computing softmax for a single vector of sequence length $l$. The results are divided into two parts: one for $l\leq1024$ and another for $l>1024$. 

As shown in Fig.\ref{fig:softmax-results-latency}, the proposed UCLM consistently achieves lower latency compared to the PUMA. While multi-core softmax support results in similar latency at smaller $l$, the improvements become more pronounced at longer $l$. For instance, at $l=8192$ and ALU width of $16$, the latency of softmax for the PUMA reaches $22.13\mu s$. The proposed UCLM lowers this to $6 \mu s$, and with multi-core softmax support, it is further reduced to $1.36 \mu s$. At smaller $l$, the parallelism of \verb|exp| from UCLMs within a core is sufficient to compute softmax, and thus no difference is observed with multi-core softmax support.

In addition to multi-core support, the performance of softmax can also be enhanced by increasing the ALU width. A wider ALU width enables greater parallelism for element-wise operations such as \verb|max|, \verb|sub|, \verb|add|, and \verb|div|, which are essential components of the softmax computation. For instance, at a sequence length of $8192$, increasing ALU width from $16$ to $64$ improves the softmax computation performance by 22\% in the multi-core configuration. 

\begin{figure}[t]
    \centering
    \includegraphics[width=\columnwidth, trim={0 5mm 0 0 0}]{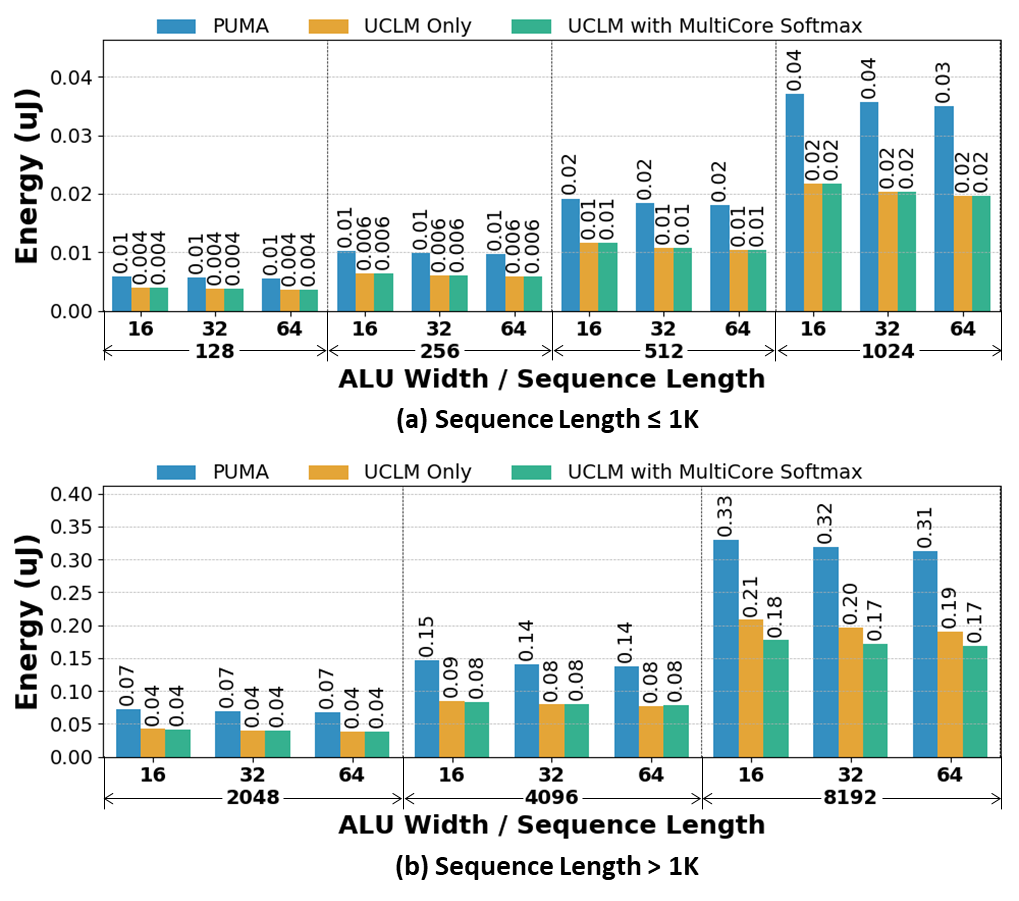}
    \caption{Energy consumption of the proposed UCLM-based hardware and multi-core softmax computation support compared to PUMA.}
    \label{fig:softmax-results-energy}
\end{figure}

Energy consumption follows similar trends, as shown in Fig.\ref{fig:softmax-results-energy}, with the PUMA consuming the most energy that increases with the sequence length. 
For $l \leq 1024$, softmax computation is extremely cheap for all configurations. 
While the difference between PUMA and our architecture increases for $l>1024$, the ratio remains same at approximately $1.6\times$. The small energy difference between the UCLM only and the multi-core softmax support demonstrates the effectiveness of gathering maxima/partial sums in improving performance.

\subsection{Transformer Layer Acceleration}
\begin{figure}[t!]
    \centering
    \includegraphics[width=\columnwidth, trim={0 5mm 0 0 0}]{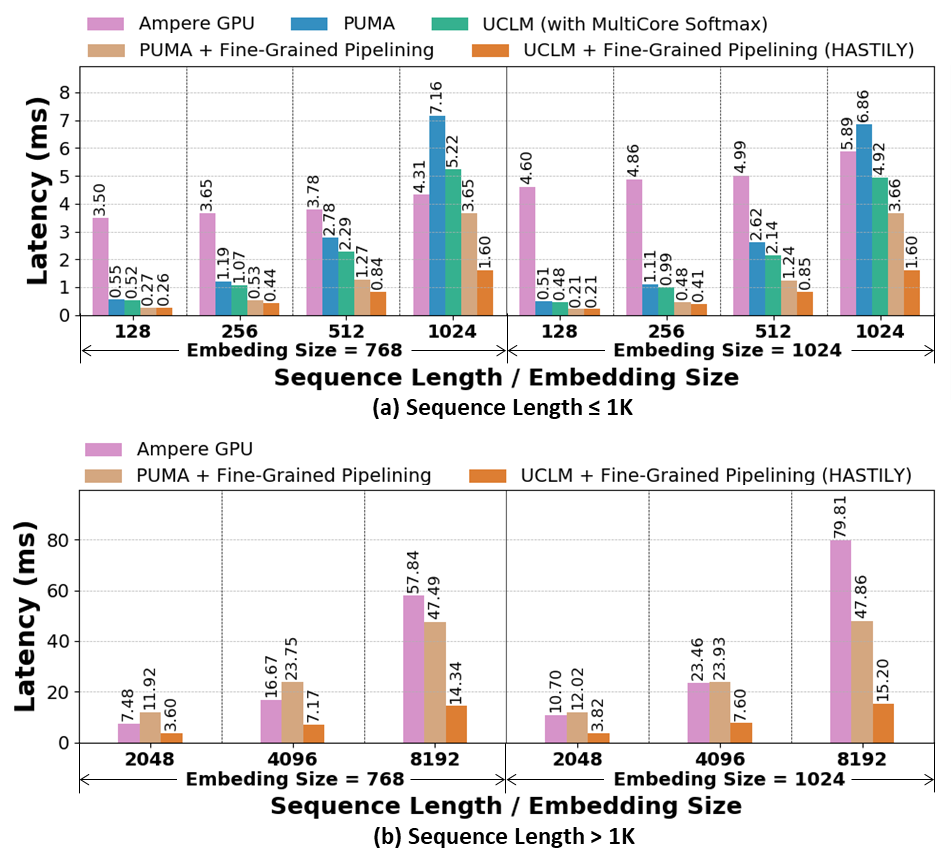}
    \caption{Compute latency of an encoder layer across varying sequence length and embedding sizes, showing the benefits of softmax acceleration and fine-grained pipelining compared to PUMA and GPU.}
    \label{fig:layer-results-latency}
\end{figure}

\begin{figure}[t]
    \centering
    \includegraphics[width=\columnwidth, trim={0 10mm 0 0 0}]{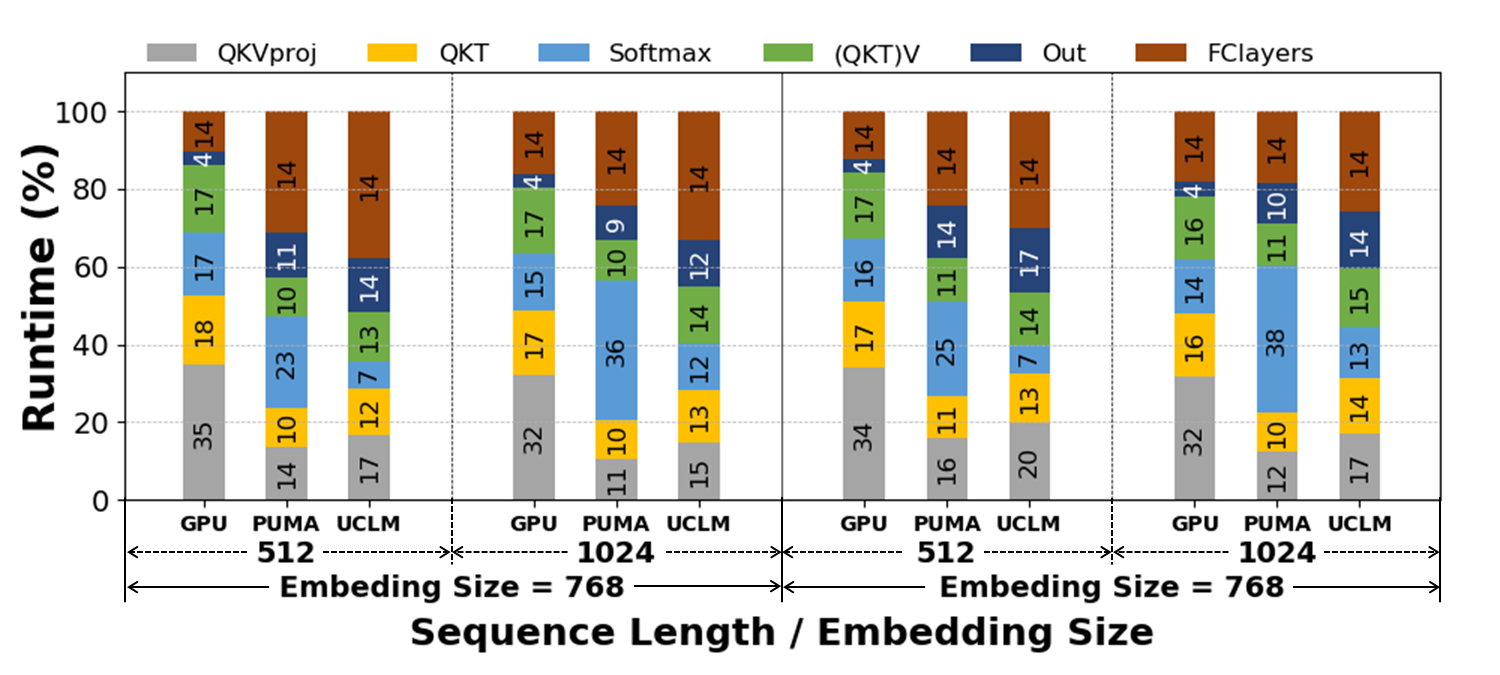}
    \caption{Runtime distribution of compute blocks (listed in legend) in an encoder layer for embedding size of 768 and 1024 and sequence length of 512 and 1024. The x-axis shows different hardware including the Ampere GPU, PUMA, and the UCLM with multicore softmax. }
    \label{fig:runtime-distribution}
\end{figure}

\begin{figure}[t!]
    \centering
    \includegraphics[width=\columnwidth, trim={0 6mm 0 0 0}]{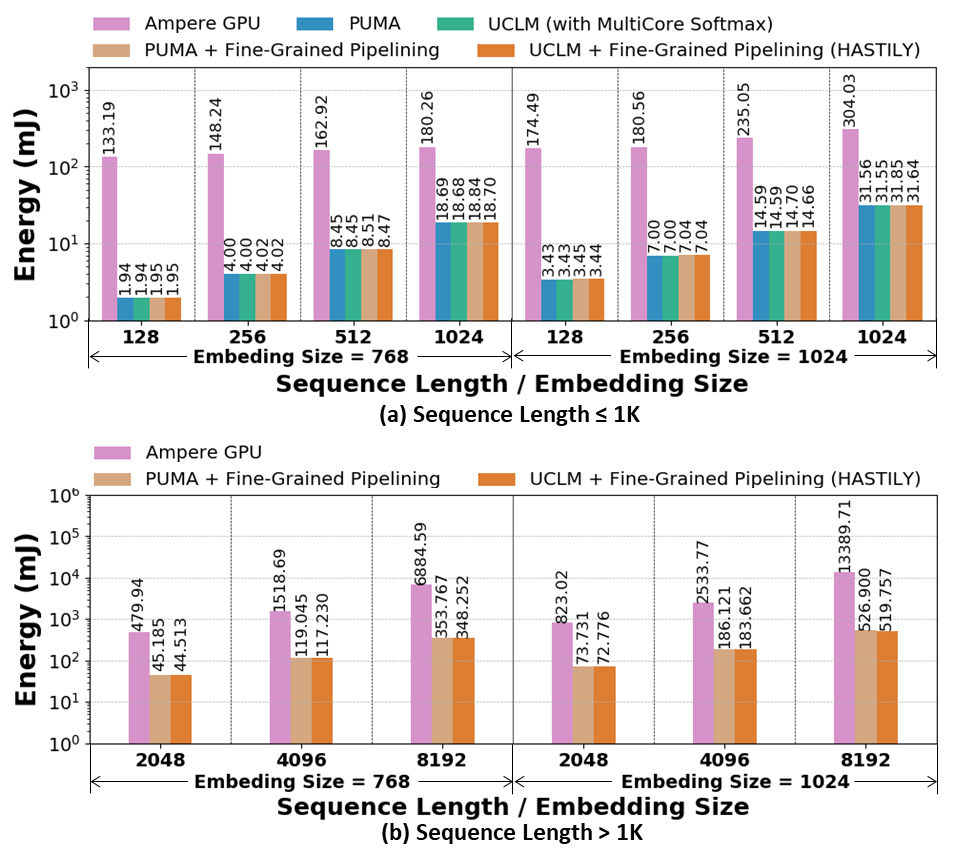}
    \caption{Energy consumption of an encoder layer across varying sequence length and embedding sizes, showing the benefits of softmax acceleration and fine-grained pipelining compared to PUMA and GPU.}
    \label{fig:layer-results-energy}
\end{figure}

The evaluations for a full encoder layer, including both attention and feed-forward blocks, are shown in Fig.\ref{fig:layer-results-latency}, \ref{fig:runtime-distribution}, and \ref{fig:layer-results-energy}. 
These results consider varying sequence length and embedding sizes, with a head size of $64$.
The figure highlights the impact of softmax acceleration and the fine-grained pipelining. 
For $l>1024$, only the results for GPUs and CIM cores with fine-grained pipelining are shown, as the chip is unable to accommodate the intermediate matrices that grow quadratically with the sequence length.

While the PUMA outperforms the GPU at $l < 1024$, it becomes slower at 1024. For $l = 2048$ and $4096$, even the PUMA with finer-granularity pipelining is worse than GPU, though it regains superiority at $8192$. The proposed UCLM with multicore softmax helps in significantly reducing the softmax computation and achieves better results than the PUMA. The effectiveness of softmax acceleration grows with increasing sequence length. At a sequence length of $8192$, the proposed architecture outperforms the PUMA with fine-grained pipelining by $3.31\times$.

At $l \leq 1024$, finer-granularity pipelining proves more effective than softmax acceleration alone, and their combination yields even greater performance gains. For instance, with an embedding size of $768$ and $l=1024$, softmax acceleration improves speed by 37\%, finer-granularity pipelining by 96\%, and their combined effect achieves a $4.47\times$ speedup over the PUMA.

Furthermore, regardless of sequence length, the proposed architecture consistently outperforms the GPU. The integration of fine-grained pipelining with our core architecture further enhances the performance, $3\times$ $-13\times$ faster than GPU. The latency with fine-grained pipelining scales linearly with the sequence length as expected from the hypothesis. Note again, smaller sequence lengths greatly benefit with all CIM architectures. At higher sequence lengths, the latency benefits of our architecture, at almost no area cost, are more evident.

Fig. \ref{fig:runtime-distribution} contrasts the runtime distribution of compute blocks, including $Q/K/V$ projection, $QK^T$, softmax, $QK^TV$, multi-head output transformation and fully connected layers, across the proposed UCLM-based architecture, the Ampere GPU, and PUMA. 
Note, the results exclude fine-grained pipelining since runtime cannot be accurately measured when these compute blocks are pipelined.
As shown in  Fig. \ref{fig:layer-results-latency}, PUMA exhibits higher latency than the GPU for $l=1024$. The runtime breakdown reveals that this higher latency in PUMA is primarily due to its high softmax computation time, which accounts for 38\% of the total runtime. 
The graph demonstrates that the UCLM-based architecture significantly reduces the runtime percentage spent on softmax in an encoder layer. For instance, at $l=1024$, PUMA's 38\% runtime spent on softmax is reduced to 13\% in UCLM-based architecture with multicore support.

For the dynamic energy (Fig. \ref{fig:layer-results-energy}), our work achieves significantly lower energy consumption compared to GPU, ranging from $10\times$ $-68\times$ reduction. 
Between PUMA and the proposed architectures, the difference in energy consumption is negligible. This is because the instructions and the number of internal device accesses within the chip remain nearly identical. Additionally, ADC energy dominates the overall energy consumption, further minimizing the impact of architectural differences on energy usage.

\begin{figure*}[t]
    \centering
    \includegraphics[width=\textwidth, trim={0 5mm 0 0 0}]{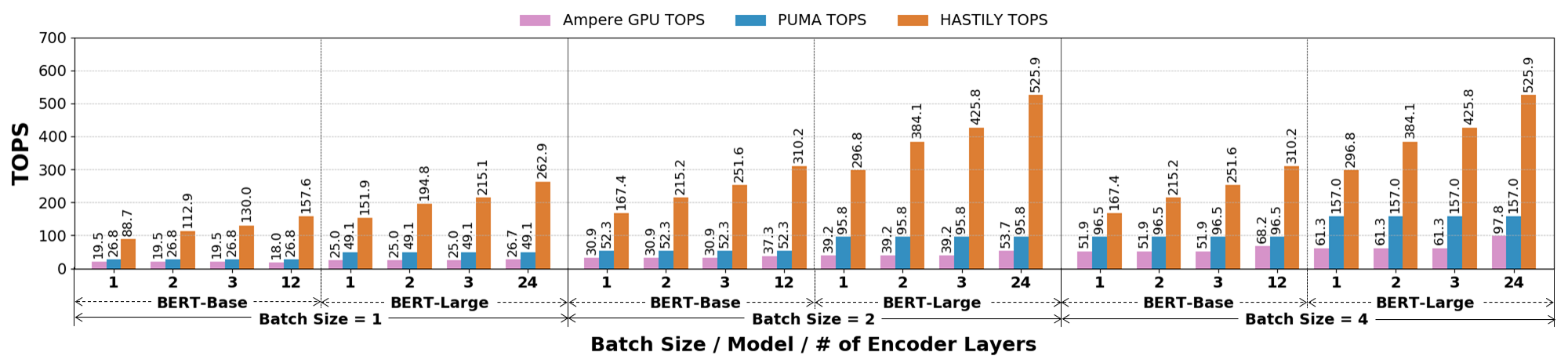}
    \caption{Comparison of our accelerator compared to PUMA and GPU in terms of effective $TOPS$ for multiple encoder layers of BERT-Base and BERT-Large models. $TOPS$ is calculated by adding the total multiplication and addition operations from matrix-multiplications involved in transformers.}
    \label{fig:end2end-results-latency}
\end{figure*}

\begin{figure*}[t]
    \centering
    \includegraphics[width=\textwidth, trim={0 5mm 0 0 0}]{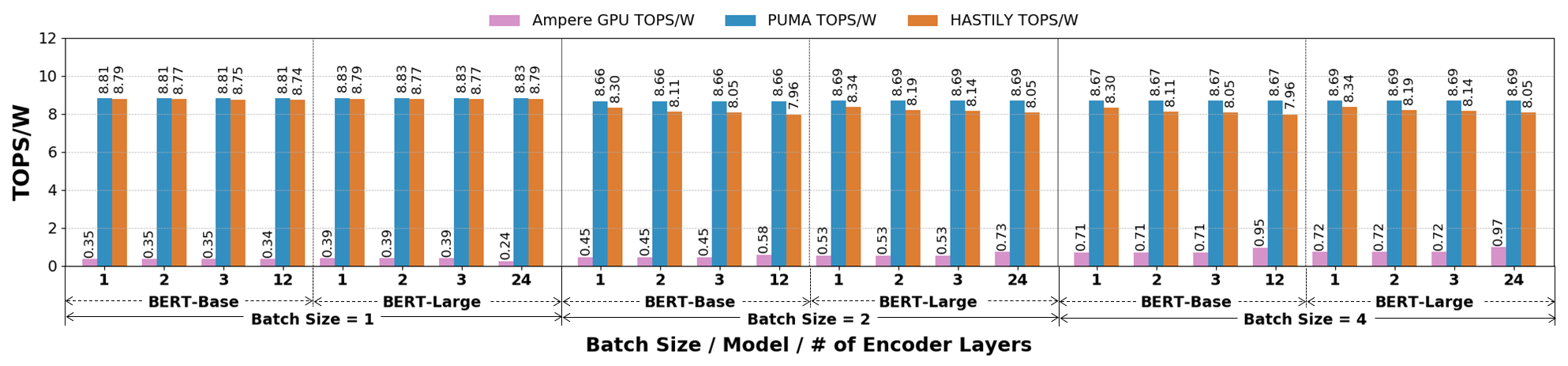}
    \caption{Comparison of our accelerator compared to PUMA and GPU in terms of effective $TOPS/W$ for multiple encoder layers of BERT-Base and BERT-Large models. }
    \label{fig:end2end-results-energy}
\end{figure*}

\subsection{End-to-End performance and energy efficiency}
To comprehensively evaluate the effectiveness and scalability of our approach, we analyzed its performance across different numbers of encoder layers in BERT models and batch size, comparing it with the Ampere GPU and PUMA. The metrics assessed include compute efficiency ($TOPS$) and energy efficiency ($TOPS/W$), where $TOPS$ is tera operations per second and the number of operations is the sum of multiplication and addition operations resulting from all the matrix-multiplication computations in the given layers.

The compute efficiency results, depicted in Fig. \ref{fig:end2end-results-latency}(a), show the superior performance of our accelerator compared to both the Ampere GPU and PUMA. With a batch size of 1, the GPU and PUMA deliver similar performance across various encoder layer configurations within a given model. For instance, the GPU achieves $19~ TOPS$, while the PUMA reaches $26~ TOPS$ for the BERT-Base model. Both are clearly under-utilized for BERT models with a single batch size. In contrast, our architecture shows improved performance as the number of encoder layers increases. This is attributed to its ability to hide the latency of computation blocks by overlapping it with other computations, as illustrated in Fig. \ref{fig:fine-gran-timeline}(b). 

In CIM architectures, utilizing multiple batch sizes can be achieved either by replicating model weight parameters across additional hardware for parallel computation or by processing batches serially. In our hardware configuration, PUMA can store weights for up to four batches within a single layer, leading to increased throughput as the batch size grows. However, our proposed fine-grained pipelining requires sufficient memory to store weights for at least two layers. This allows both BERT-Base and BERT-Large models to support finer-grained processing for batch sizes up to 2. For batch size 4, computations for batch size 2 are repeated, resulting in performance identical to that of batch size 2.
Overall, our accelerator achieves $158~ TOPS$ and $263~ TOPS$ for the BERT-Base and BERT-Large models, respectively, outperforming the GPU by up to $9.8\times$.

Energy efficiency, as illustrated in Fig.\ref{fig:end2end-results-energy}, follows the trend observed for a single encoder layer. 
The Ampere GPU exhibits low energy efficiency due to under-utilization, achieving only $0.3~TOPS/W$ for a single batch. Its efficiency improves with increasing batch size, reaching up to $0.9~TOPS/W$.
In contrast, HASTILY consistently achieves high energy efficiency of $8~TOPS/W$, regardless of the model size, number of encoder layers, or batch size. These results highlight the superior scalability and performance of our CIM-based architecture.

\section{Discussion and Future Work}
\label{sec:discussion}

\textbf{Scope:} Our results focus on highlighting the benefits of HASTILY for BERT models, with bigger improvements at longer sequence lengths. Since vision transformers also have an encoder-like architecture, the findings are also applicable to them \cite{dosovitskiy2021imageworth16x16words}. For decoder-like models such as GPT, our accelerator will give similar improvements for the prefill-phase. However, the decoder-phase of GPT models is dominantly memory-bound and the benefits of softmax acceleration will be limited for such use cases. 

\textbf{Accuracy:} Regarding accuracy, we primarily utilize low-precision (8-bit) for our evaluations, owing to the ongoing trend of using quantized models for inference \cite{dettmers2022llm, kim2021bert}. Further, analog CIM also introduces quantization and functional errors due to various circuit non-idealities. We assume that such errors can be reduced during transformer inference using hardware-aware training \cite{joshi2020accurate}. 

\textbf{Future work:} HASTILY uses the dual-functionality SRAM arrays for storing lookup tables for exponent operations. It can be also extended to accommodate pre-computed tables for other transcendental functions, including $log$ and $sin$ \cite{intelHarrison1999TheCO}. Our evaluation setup allows exploring other hardware configurations, model parameters and quantization, but it can be incorporated with digital CIM to expand the design space in future. Additionally, though our softmax implementation is computationally stable because we take care of the overflow/underflow conditions, it can be modified to include the logsoftmax or online implementations if needed \cite{blanchard2021accurately,dao2022flashattention}. Integrating our fine-granularity pipelining technique in GPUs could also be an interesting study to further investigate the benefits of our proposed optimization.

\section{Conclusion}
\label{sec:conclusion}
In this paper, we introduced a scalable area-efficient compute-in-memory (CIM) accelerator, HASTILY, with unified compute and lookup modules (UCLMs) to accelerate transformer models. Leveraging parallelism at multiple levels, including UCLMs, cores and tiles, our spatial architecture efficiently computes both softmax and matrix-multiplication, solving the softmax bottleneck. The evaluations demonstrated significant contributions from both the UCLM based hardware and the software optimizations such as multi-core softmax support and fine-granularity pipelining in improving the final throughput for BERT models on CIM accelerator. Finally, we believe the compiler and cycle-level simulator tailored towards transformer inference on CIM architecture would open up new research directions, taking the field forward.

\section*{Aknowledgments}
This work was supported in part by the Center for the Co-Design of Cognitive Systems (COCOSYS), in part by the SRC/DARPA JUMP centers.

\bibliographystyle{IEEEtran}
\bibliography{refs}

\begin{thebibliography}{10}
\providecommand{\url}[1]{#1}
\csname url@samestyle\endcsname
\providecommand{\newblock}{\relax}
\providecommand{\bibinfo}[2]{#2}
\providecommand{\BIBentrySTDinterwordspacing}{\spaceskip=0pt\relax}
\providecommand{\BIBentryALTinterwordstretchfactor}{4}
\providecommand{\BIBentryALTinterwordspacing}{\spaceskip=\fontdimen2\font plus
\BIBentryALTinterwordstretchfactor\fontdimen3\font minus \fontdimen4\font\relax}
\providecommand{\BIBforeignlanguage}[2]{{%
\expandafter\ifx\csname l@#1\endcsname\relax
\typeout{** WARNING: IEEEtran.bst: No hyphenation pattern has been}%
\typeout{** loaded for the language `#1'. Using the pattern for}%
\typeout{** the default language instead.}%
\else
\language=\csname l@#1\endcsname
\fi
#2}}
\providecommand{\BIBdecl}{\relax}
\BIBdecl

\bibitem{AttentionIsAllYouNeed}
A.~Vaswani, N.~Shazeer, N.~Parmar, J.~Uszkoreit, L.~Jones, A.~N. Gomez, L.~Kaiser, and I.~Polosukhin, ``Attention is all you need,'' in \emph{Proceedings of the 31st International Conference on Neural Information Processing Systems}, ser. NIPS'17.\hskip 1em plus 0.5em minus 0.4em\relax Red Hook, NY, USA: Curran Associates Inc., 2017, p. 6000–6010.

\bibitem{Ampere}
NVIDIA, ``Nvidia ampere ga-102 gpu architecture,'' \url{https://www.nvidia.com/content/PDF/nvidia-ampere-ga-102-gpu-architecture-whitepaper-v2.pdf}.

\bibitem{conneau2019cross}
A.~Conneau and G.~Lample, ``Cross-lingual language model pretraining,'' \emph{Advances in neural information processing systems}, vol.~32, 2019.

\bibitem{dai2019transformer}
Z.~Dai, ``Transformer-xl: Attentive language models beyond a fixed-length context,'' \emph{arXiv preprint arXiv:1901.02860}, 2019.

\bibitem{swinTransformer}
\BIBentryALTinterwordspacing
Z.~Liu, Y.~Lin, Y.~Cao, H.~Hu, Y.~Wei, Z.~Zhang, S.~Lin, and B.~Guo, ``Swin transformer: Hierarchical vision transformer using shifted windows,'' 2021. [Online]. Available: \url{https://arxiv.org/abs/2103.14030}
\BIBentrySTDinterwordspacing

\bibitem{dosovitskiy2021imageworth16x16words}
\BIBentryALTinterwordspacing
A.~Dosovitskiy, L.~Beyer, A.~Kolesnikov, D.~Weissenborn, X.~Zhai, T.~Unterthiner, M.~Dehghani, M.~Minderer, G.~Heigold, S.~Gelly, J.~Uszkoreit, and N.~Houlsby, ``An image is worth 16x16 words: Transformers for image recognition at scale,'' 2021. [Online]. Available: \url{https://arxiv.org/abs/2010.11929}
\BIBentrySTDinterwordspacing

\bibitem{bello2020attentionaugmentedconvolutionalnetworks}
\BIBentryALTinterwordspacing
I.~Bello, B.~Zoph, A.~Vaswani, J.~Shlens, and Q.~V. Le, ``Attention augmented convolutional networks,'' 2020. [Online]. Available: \url{https://arxiv.org/abs/1904.09925}
\BIBentrySTDinterwordspacing

\bibitem{chen2020generative}
M.~Chen, A.~Radford, R.~Child, J.~Wu, H.~Jun, D.~Luan, and I.~Sutskever, ``Generative pretraining from pixels,'' in \emph{International conference on machine learning}.\hskip 1em plus 0.5em minus 0.4em\relax PMLR, 2020, pp. 1691--1703.

\bibitem{wu2021cvt}
H.~Wu, B.~Xiao, N.~Codella, M.~Liu, X.~Dai, L.~Yuan, and L.~Zhang, ``Cvt: Introducing convolutions to vision transformers,'' in \emph{Proceedings of the IEEE/CVF international conference on computer vision}, 2021, pp. 22--31.

\bibitem{sun2020transtrack}
P.~Sun, J.~Cao, Y.~Jiang, R.~Zhang, E.~Xie, Z.~Yuan, C.~Wang, and P.~Luo, ``Transtrack: Multiple object tracking with transformer,'' \emph{arXiv preprint arXiv:2012.15460}, 2020.

\bibitem{A3}
\BIBentryALTinterwordspacing
T.~J. Ham, S.~Jung, S.~Kim, Y.~H. Oh, Y.~Park, Y.~Song, J.~Park, S.~Lee, K.~Park, J.~W. Lee, and D.~Jeong, ``A\({}^{\mbox{3}}\): Accelerating attention mechanisms in neural networks with approximation,'' \emph{CoRR}, vol. abs/2002.10941, 2020. [Online]. Available: \url{https://arxiv.org/abs/2002.10941}
\BIBentrySTDinterwordspacing

\bibitem{Spatten}
\BIBentryALTinterwordspacing
H.~Wang, Z.~Zhang, and S.~Han, ``Spatten: Efficient sparse attention architecture with cascade token and head pruning,'' \emph{CoRR}, vol. abs/2012.09852, 2020. [Online]. Available: \url{https://arxiv.org/abs/2012.09852}
\BIBentrySTDinterwordspacing

\bibitem{EdgeBERT}
\BIBentryALTinterwordspacing
T.~Tambe, C.~Hooper, L.~Pentecost, T.~Jia, E.-Y. Yang, M.~Donato, V.~Sanh, P.~Whatmough, A.~M. Rush, D.~Brooks, and G.-Y. Wei, ``Edgebert: Sentence-level energy optimizations for latency-aware multi-task nlp inference,'' in \emph{MICRO-54: 54th Annual IEEE/ACM International Symposium on Microarchitecture}, ser. MICRO '21.\hskip 1em plus 0.5em minus 0.4em\relax New York, NY, USA: Association for Computing Machinery, 2021, p. 830–844. [Online]. Available: \url{https://doi.org/10.1145/3466752.3480095}
\BIBentrySTDinterwordspacing

\bibitem{lu2021sanger}
L.~Lu, Y.~Jin, H.~Bi, Z.~Luo, P.~Li, T.~Wang, and Y.~Liang, ``Sanger: A co-design framework for enabling sparse attention using reconfigurable architecture,'' in \emph{MICRO-54: 54th Annual IEEE/ACM International Symposium on Microarchitecture}, 2021, pp. 977--991.

\bibitem{softermax}
\BIBentryALTinterwordspacing
J.~R. Stevens, R.~Venkatesan, S.~Dai, B.~Khailany, and A.~Raghunathan, ``Softermax: Hardware/software co-design of an efficient softmax for transformers,'' in \emph{2021 58th ACM/IEEE Design Automation Conference (DAC)}.\hskip 1em plus 0.5em minus 0.4em\relax IEEE Press, 2021, p. 469–474. [Online]. Available: \url{https://doi.org/10.1109/DAC18074.2021.9586134}
\BIBentrySTDinterwordspacing

\bibitem{dao2022flashattention}
T.~Dao, D.~Fu, S.~Ermon, A.~Rudra, and C.~R{\'e}, ``Flashattention: Fast and memory-efficient exact attention with io-awareness,'' \emph{Advances in Neural Information Processing Systems}, vol.~35, pp. 16\,344--16\,359, 2022.

\bibitem{kao2023flat}
S.-C. Kao, S.~Subramanian, G.~Agrawal, A.~Yazdanbakhsh, and T.~Krishna, ``Flat: An optimized dataflow for mitigating attention bottlenecks,'' in \emph{Proceedings of the 28th ACM International Conference on Architectural Support for Programming Languages and Operating Systems, Volume 2}, 2023, pp. 295--310.

\bibitem{kwon2023efficient}
W.~Kwon, Z.~Li, S.~Zhuang, Y.~Sheng, L.~Zheng, C.~H. Yu, J.~Gonzalez, H.~Zhang, and I.~Stoica, ``Efficient memory management for large language model serving with pagedattention,'' in \emph{Proceedings of the 29th Symposium on Operating Systems Principles}, 2023, pp. 611--626.

\bibitem{zadeh2020gobo}
A.~H. Zadeh, I.~Edo, O.~M. Awad, and A.~Moshovos, ``Gobo: Quantizing attention-based nlp models for low latency and energy efficient inference,'' in \emph{2020 53rd Annual IEEE/ACM International Symposium on Microarchitecture (MICRO)}.\hskip 1em plus 0.5em minus 0.4em\relax IEEE, 2020, pp. 811--824.

\bibitem{softmax1_hyft}
\BIBentryALTinterwordspacing
T.~Xia and S.~Q. Zhang, ``Hyft: A reconfigurable softmax accelerator with hybrid numeric format for both training and inference,'' in \emph{Proceedings of the 29th ACM/IEEE International Symposium on Low Power Electronics and Design}, ser. ISLPED '24.\hskip 1em plus 0.5em minus 0.4em\relax New York, NY, USA: Association for Computing Machinery, 2024, p. 1–6. [Online]. Available: \url{https://doi.org/10.1145/3665314.3670816}
\BIBentrySTDinterwordspacing

\bibitem{softmax2_ITA}
\BIBentryALTinterwordspacing
G.~Islamoglu, M.~Scherer, G.~Paulin, T.~Fischer, V.~J.~B. Jung, A.~Garofalo, and L.~Benini, ``Ita: An energy-efficient attention and softmax accelerator for quantized transformers.'' in \emph{ISLPED}.\hskip 1em plus 0.5em minus 0.4em\relax IEEE, 2023, pp. 1--6. [Online]. Available: \url{http://dblp.uni-trier.de/db/conf/islped/islped2023.html#IslamogluSPFJGB23}
\BIBentrySTDinterwordspacing

\bibitem{softmax3}
N.~A. Koca, A.~T. Do, and C.-H. Chang, ``Hardware-efficient softmax approximation for self-attention networks,'' in \emph{2023 IEEE International Symposium on Circuits and Systems (ISCAS)}, 2023, pp. 1--5.

\bibitem{dai-etal-2019-transformer}
\BIBentryALTinterwordspacing
Z.~Dai, Z.~Yang, Y.~Yang, J.~Carbonell, Q.~Le, and R.~Salakhutdinov, ``Transformer-{XL}: Attentive language models beyond a fixed-length context,'' in \emph{Proceedings of the 57th Annual Meeting of the Association for Computational Linguistics}, A.~Korhonen, D.~Traum, and L.~M{\`a}rquez, Eds.\hskip 1em plus 0.5em minus 0.4em\relax Florence, Italy: Association for Computational Linguistics, Jul. 2019, pp. 2978--2988. [Online]. Available: \url{https://aclanthology.org/P19-1285}
\BIBentrySTDinterwordspacing

\bibitem{choromanski2022rethinkingattentionperformers}
\BIBentryALTinterwordspacing
K.~Choromanski, V.~Likhosherstov, D.~Dohan, X.~Song, A.~Gane, T.~Sarlos, P.~Hawkins, J.~Davis, A.~Mohiuddin, L.~Kaiser, D.~Belanger, L.~Colwell, and A.~Weller, ``Rethinking attention with performers,'' 2022. [Online]. Available: \url{https://arxiv.org/abs/2009.14794}
\BIBentrySTDinterwordspacing

\bibitem{retransformer}
X.~Yang, B.~Yan, H.~Li, and Y.~Chen, ``Retransformer: Reram-based processing-in-memory architecture for transformer acceleration,'' in \emph{Proceedings of the 39th International Conference on Computer-Aided Design}, 2020, pp. 1--9.

\bibitem{xformer}
S.~Sridharan, J.~R. Stevens, K.~Roy, and A.~Raghunathan, ``X-former: In-memory acceleration of transformers,'' \emph{IEEE Transactions on Very Large Scale Integration (VLSI) Systems}, vol.~31, no.~8, pp. 1223--1233, 2023.

\bibitem{wulf1995hitting}
W.~A. Wulf and S.~A. McKee, ``Hitting the memory wall: Implications of the obvious,'' \emph{ACM SIGARCH computer architecture news}, vol.~23, no.~1, pp. 20--24, 1995.

\bibitem{gholami2024ai}
A.~Gholami, Z.~Yao, S.~Kim, C.~Hooper, M.~W. Mahoney, and K.~Keutzer, ``Ai and memory wall,'' \emph{IEEE Micro}, 2024.

\bibitem{verma2019memory}
N.~Verma, H.~Jia, H.~Valavi, Y.~Tang, M.~Ozatay, L.-Y. Chen, B.~Zhang, and P.~Deaville, ``In-memory computing: Advances and prospects,'' \emph{IEEE Solid-State Circuits Magazine}, vol.~11, no.~3, pp. 43--55, 2019.

\bibitem{Isaac}
\BIBentryALTinterwordspacing
A.~Shafiee, A.~Nag, N.~Muralimanohar, R.~Balasubramonian, J.~P. Strachan, M.~Hu, R.~S. Williams, and V.~Srikumar, ``Isaac: a convolutional neural network accelerator with in-situ analog arithmetic in crossbars,'' in \emph{Proceedings of the 43rd International Symposium on Computer Architecture}, ser. ISCA '16.\hskip 1em plus 0.5em minus 0.4em\relax IEEE Press, 2016, p. 14–26. [Online]. Available: \url{https://doi.org/10.1109/ISCA.2016.12}
\BIBentrySTDinterwordspacing

\bibitem{aliCIM}
M.~Ali, S.~Roy, U.~Saxena, T.~Sharma, A.~Raghunathan, and K.~Roy, ``Compute-in-memory technologies and architectures for deep learning workloads,'' \emph{IEEE Transactions on Very Large Scale Integration (VLSI) Systems}, vol.~30, no.~11, pp. 1615--1630, 2022.

\bibitem{imact}
A.~F. Laguna, A.~Kazemi, M.~Niemier, and X.~S. Hu, ``In-memory computing based accelerator for transformer networks for long sequences,'' in \emph{2021 Design, Automation \& Test in Europe Conference \& Exhibition (DATE)}.\hskip 1em plus 0.5em minus 0.4em\relax IEEE, 2021, pp. 1839--1844.

\bibitem{transpim}
M.~Zhou, W.~Xu, J.~Kang, and T.~Rosing, ``Transpim: A memory-based acceleration via software-hardware co-design for transformer,'' in \emph{2022 IEEE International Symposium on High-Performance Computer Architecture (HPCA)}.\hskip 1em plus 0.5em minus 0.4em\relax IEEE, 2022, pp. 1071--1085.

\bibitem{devlin2018bert}
J.~Devlin, M.-W. Chang, K.~Lee, and K.~Toutanova, ``Bert: Pre-training of deep bidirectional transformers for language understanding,'' \emph{arXiv preprint arXiv:1810.04805}, 2018.

\bibitem{sun2020finetuneberttextclassification}
\BIBentryALTinterwordspacing
C.~Sun, X.~Qiu, Y.~Xu, and X.~Huang, ``How to fine-tune bert for text classification?'' 2020. [Online]. Available: \url{https://arxiv.org/abs/1905.05583}
\BIBentrySTDinterwordspacing

\bibitem{bertTextClassification}
\BIBentryALTinterwordspacing
E.~C. Garrido-Merchan, R.~Gozalo-Brizuela, and S.~Gonzalez-Carvajal, ``Comparing bert against traditional machine learning models in text classification,'' \emph{Journal of Computational and Cognitive Engineering}, vol.~2, no.~4, p. 352–356, Apr. 2023. [Online]. Available: \url{http://dx.doi.org/10.47852/bonviewJCCE3202838}
\BIBentrySTDinterwordspacing

\bibitem{kocián2021siamesebertbasedmodelweb}
\BIBentryALTinterwordspacing
M.~Kocián, J.~Náplava, D.~Štancl, and V.~Kadlec, ``Siamese bert-based model for web search relevance ranking evaluated on a new czech dataset,'' 2021. [Online]. Available: \url{https://arxiv.org/abs/2112.01810}
\BIBentrySTDinterwordspacing

\bibitem{yenduri2023gpt}
G.~Yenduri, M.~Ramalingam, G.~Chemmalar~Selvi, Y.~Supriya, G.~Srivastava, P.~Maddikunta, G.~Deepti~Raj, R.~Jhaveri, B.~Prabadevi, W.~Wang \emph{et~al.}, ``Gpt (generative pre-trained transformer)--a comprehensive review on enabling technologies,'' \emph{Potential Applications, Emerging Challenges, and Future Directions}, 2023.

\bibitem{goyal2023newssummarizationevaluationera}
\BIBentryALTinterwordspacing
T.~Goyal, J.~J. Li, and G.~Durrett, ``News summarization and evaluation in the era of gpt-3,'' 2023. [Online]. Available: \url{https://arxiv.org/abs/2209.12356}
\BIBentrySTDinterwordspacing

\bibitem{automatedNewsSummarization}
A.~Gupta, D.~Chugh, Anjum, and R.~Katarya, ``Automated news summarization using transformers,'' in \emph{Sustainable Advanced Computing}, S.~Aurelia, S.~S. Hiremath, K.~Subramanian, and S.~K. Biswas, Eds.\hskip 1em plus 0.5em minus 0.4em\relax Singapore: Springer Singapore, 2022, pp. 249--259.

\bibitem{thoppilan2022lamdalanguagemodelsdialog}
\BIBentryALTinterwordspacing
R.~Thoppilan, D.~D. Freitas, J.~Hall, N.~Shazeer, A.~Kulshreshtha, H.-T. Cheng, A.~Jin, T.~Bos, L.~Baker, Y.~Du, Y.~Li, H.~Lee, H.~S. Zheng, A.~Ghafouri, M.~Menegali, Y.~Huang, M.~Krikun, D.~Lepikhin, J.~Qin, D.~Chen, Y.~Xu, Z.~Chen, A.~Roberts, M.~Bosma, V.~Zhao, Y.~Zhou, C.-C. Chang, I.~Krivokon, W.~Rusch, M.~Pickett, P.~Srinivasan, L.~Man, K.~Meier-Hellstern, M.~R. Morris, T.~Doshi, R.~D. Santos, T.~Duke, J.~Soraker, B.~Zevenbergen, V.~Prabhakaran, M.~Diaz, B.~Hutchinson, K.~Olson, A.~Molina, E.~Hoffman-John, J.~Lee, L.~Aroyo, R.~Rajakumar, A.~Butryna, M.~Lamm, V.~Kuzmina, J.~Fenton, A.~Cohen, R.~Bernstein, R.~Kurzweil, B.~Aguera-Arcas, C.~Cui, M.~Croak, E.~Chi, and Q.~Le, ``Lamda: Language models for dialog applications,'' 2022. [Online]. Available: \url{https://arxiv.org/abs/2201.08239}
\BIBentrySTDinterwordspacing

\bibitem{gpt-j}
B.~Wang and A.~Komatsuzaki, ``{GPT-J-6B: A 6 Billion Parameter Autoregressive Language Model},'' \url{https://github.com/kingoflolz/mesh-transformer-jax}, May 2021.

\bibitem{zhang2022optopenpretrainedtransformer}
\BIBentryALTinterwordspacing
S.~Zhang, S.~Roller, N.~Goyal, M.~Artetxe, M.~Chen, S.~Chen, C.~Dewan, M.~Diab, X.~Li, X.~V. Lin, T.~Mihaylov, M.~Ott, S.~Shleifer, K.~Shuster, D.~Simig, P.~S. Koura, A.~Sridhar, T.~Wang, and L.~Zettlemoyer, ``Opt: Open pre-trained transformer language models,'' 2022. [Online]. Available: \url{https://arxiv.org/abs/2205.01068}
\BIBentrySTDinterwordspacing

\bibitem{you2023vitcod}
H.~You, Z.~Sun, H.~Shi, Z.~Yu, Y.~Zhao, Y.~Zhang, C.~Li, B.~Li, and Y.~Lin, ``Vitcod: Vision transformer acceleration via dedicated algorithm and accelerator co-design,'' in \emph{2023 IEEE International Symposium on High-Performance Computer Architecture (HPCA)}.\hskip 1em plus 0.5em minus 0.4em\relax IEEE, 2023, pp. 273--286.

\bibitem{zhu2023vlgptgenerativepretrainedtransformer}
\BIBentryALTinterwordspacing
J.~Zhu, X.~Ding, Y.~Ge, Y.~Ge, S.~Zhao, H.~Zhao, X.~Wang, and Y.~Shan, ``Vl-gpt: A generative pre-trained transformer for vision and language understanding and generation,'' 2023. [Online]. Available: \url{https://arxiv.org/abs/2312.09251}
\BIBentrySTDinterwordspacing

\bibitem{analog1cim}
X.~Si and et~al., ``A local computing cell and 6t sram-based computing-in-memory macro with 8-b mac operation for edge ai chips,'' \emph{IEEE Journal of Solid-State Circuits}, vol.~56, no.~9, pp. 2817--2831, 2021.

\bibitem{twin8t}
X.~Si, J.-J. Chen, Y.-N. Tu, W.-H. Huang, J.-H. Wang, Y.-C. Chiu, W.-C. Wei, S.-Y. Wu, X.~Sun, R.~Liu, S.~Yu, R.-S. Liu, C.-C. Hsieh, K.-T. Tang, Q.~Li, and M.-F. Chang, ``A twin-8t sram computation-in-memory unit-macro for multibit cnn-based ai edge processors,'' \emph{IEEE Journal of Solid-State Circuits}, vol.~55, no.~1, pp. 189--202, 2020.

\bibitem{computesram}
J.~Wang, X.~Wang, C.~Eckert, A.~Subramaniyan, R.~Das, D.~Blaauw, and D.~Sylvester, ``A 28-nm compute sram with bit-serial logic/arithmetic operations for programmable in-memory vector computing,'' \emph{IEEE Journal of Solid-State Circuits}, vol.~55, no.~1, pp. 76--86, 2020.

\bibitem{tmsc2023digital}
H.~Mori \emph{et~al.}, ``A 4nm 6163-tops/w/b $\mathbf{4790-TOPS/mm^{2}/b}$ sram based digital-computing-in-memory macro supporting bit-width flexibility and simultaneous mac and weight update,'' in \emph{2023 IEEE International Solid- State Circuits Conference (ISSCC)}, 2023, pp. 132--134.

\bibitem{chakraborty2020resistive}
I.~Chakraborty, M.~Ali, A.~Ankit, S.~Jain, S.~Roy, S.~Sridharan, A.~Agrawal, A.~Raghunathan, and K.~Roy, ``Resistive crossbars as approximate hardware building blocks for machine learning: Opportunities and challenges,'' \emph{Proceedings of the IEEE}, vol. 108, no.~12, pp. 2276--2310, 2020.

\bibitem{prime}
P.~Chi, S.~Li, C.~Xu, T.~Zhang, J.~Zhao, Y.~Liu, Y.~Wang, and Y.~Xie, ``Prime: A novel processing-in-memory architecture for neural network computation in reram-based main memory,'' in \emph{2016 ACM/IEEE 43rd Annual International Symposium on Computer Architecture (ISCA)}, 2016, pp. 27--39.

\bibitem{puma}
\BIBentryALTinterwordspacing
A.~Ankit, I.~E. Hajj, S.~R. Chalamalasetti, G.~Ndu, M.~Foltin, R.~S. Williams, P.~Faraboschi, W.-m.~W. Hwu, J.~P. Strachan, K.~Roy, and D.~S. Milojicic, ``Puma: A programmable ultra-efficient memristor-based accelerator for machine learning inference,'' in \emph{Proceedings of the Twenty-Fourth International Conference on Architectural Support for Programming Languages and Operating Systems}, ser. ASPLOS '19.\hskip 1em plus 0.5em minus 0.4em\relax New York, NY, USA: Association for Computing Machinery, 2019, p. 715–731. [Online]. Available: \url{https://doi.org/10.1145/3297858.3304049}
\BIBentrySTDinterwordspacing

\bibitem{song2017pipelayer}
L.~Song, X.~Qian, H.~Li, and Y.~Chen, ``Pipelayer: A pipelined reram-based accelerator for deep learning,'' in \emph{2017 IEEE international symposium on high performance computer architecture (HPCA)}.\hskip 1em plus 0.5em minus 0.4em\relax IEEE, 2017, pp. 541--552.

\bibitem{agrawal2021magnetoresistive}
A.~Agrawal, C.~Wang, T.~Sharma, and K.~Roy, ``Magnetoresistive circuits and systems: Embedded non-volatile memory to crossbar arrays,'' \emph{IEEE Transactions on Circuits and Systems I: Regular Papers}, vol.~68, no.~6, pp. 2281--2294, 2021.

\bibitem{fong2015embedding}
X.~Fong, R.~Venkatesan, D.~Lee, A.~Raghunathan, and K.~Roy, ``Embedding read-only memory in spin-transfer torque mram-based on-chip caches,'' \emph{IEEE Transactions on Very Large Scale Integration (VLSI) Systems}, vol.~24, no.~3, pp. 992--1002, 2015.

\bibitem{lee2012area}
D.~Lee and K.~Roy, ``Area efficient rom-embedded sram cache,'' \emph{IEEE Transactions on Very Large Scale Integration (VLSI) Systems}, vol.~21, no.~9, pp. 1583--1595, 2012.

\bibitem{intelHarrison1999TheCO}
\BIBentryALTinterwordspacing
J.~Harrison, P.~Tak, and P.~Tang, ``The computation of transcendental functions on the ia-64 architecture,'' 1999. [Online]. Available: \url{https://api.semanticscholar.org/CorpusID:16091395}
\BIBentrySTDinterwordspacing

\bibitem{kim2021bert}
S.~Kim, A.~Gholami, Z.~Yao, M.~W. Mahoney, and K.~Keutzer, ``I-bert: Integer-only bert quantization,'' in \emph{International conference on machine learning}.\hskip 1em plus 0.5em minus 0.4em\relax PMLR, 2021, pp. 5506--5518.

\bibitem{dettmers2022llm}
T.~Dettmers, M.~Lewis, Y.~Belkada, and L.~Zettlemoyer, ``Llm. int8 (): 8-bit matrix multiplication for transformers at scale,'' \emph{arXiv preprint arXiv:2208.07339}, 2022.

\bibitem{joshi2020accurate}
V.~Joshi, M.~Le~Gallo, S.~Haefeli, I.~Boybat, S.~R. Nandakumar, C.~Piveteau, M.~Dazzi, B.~Rajendran, A.~Sebastian, and E.~Eleftheriou, ``Accurate deep neural network inference using computational phase-change memory,'' \emph{Nature communications}, vol.~11, no.~1, p. 2473, 2020.

\bibitem{blanchard2021accurately}
P.~Blanchard, D.~J. Higham, and N.~J. Higham, ``Accurately computing the log-sum-exp and softmax functions,'' \emph{IMA Journal of Numerical Analysis}, vol.~41, no.~4, pp. 2311--2330, 2021.

\end{thebibliography}

\begin{IEEEbiography}[{\includegraphics[width=1in,height=1.25in,clip,keepaspectratio]{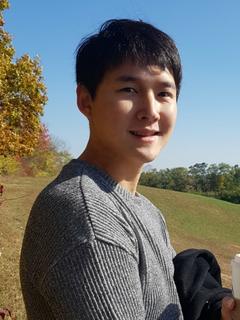}}]{Dong Eun Kim} 
received his B.S. in Electrical and Electronic Engineering from Yonsei University, Korea, in 2018. Currently, he is pursuing a Ph.D. degree in Electrical and Computer Engineering at Purdue University under the supervision of Professor Kaushik Roy. His research interests lie in hardware and software do-design for neural network. Particularly he is interested in in-memory computing based on non-volatile memories. During his PhD, he has done internship with IBM in 2021. 
\end{IEEEbiography}

\begin{IEEEbiography}[{\includegraphics[width=1in,height=1.25in,clip,keepaspectratio]{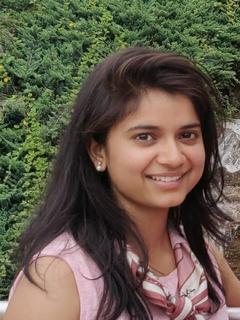}}]{Tanvi Sharma} received her bachelor’s degree from Indian Institute of Technology, Roorkee in 2018 and worked as a Digital Design Engineer at Texas Instruments before joining Purdue University in 2019. She is in the direct PhD program at Purdue, under the guidance of Professor Kaushik Roy. She has been a recipient of Qualcomn Innovation Fellowship in 2021, and MLSys Rising Stars Award in 2023. Her research interests lie at the intersection of machine learning and systems, with focus on developing energy efficient ML hardware accelerators using compute-in-memory solutions.
\end{IEEEbiography}

\begin{IEEEbiography}[{\includegraphics[width=1in,height=1.25in,clip,keepaspectratio]{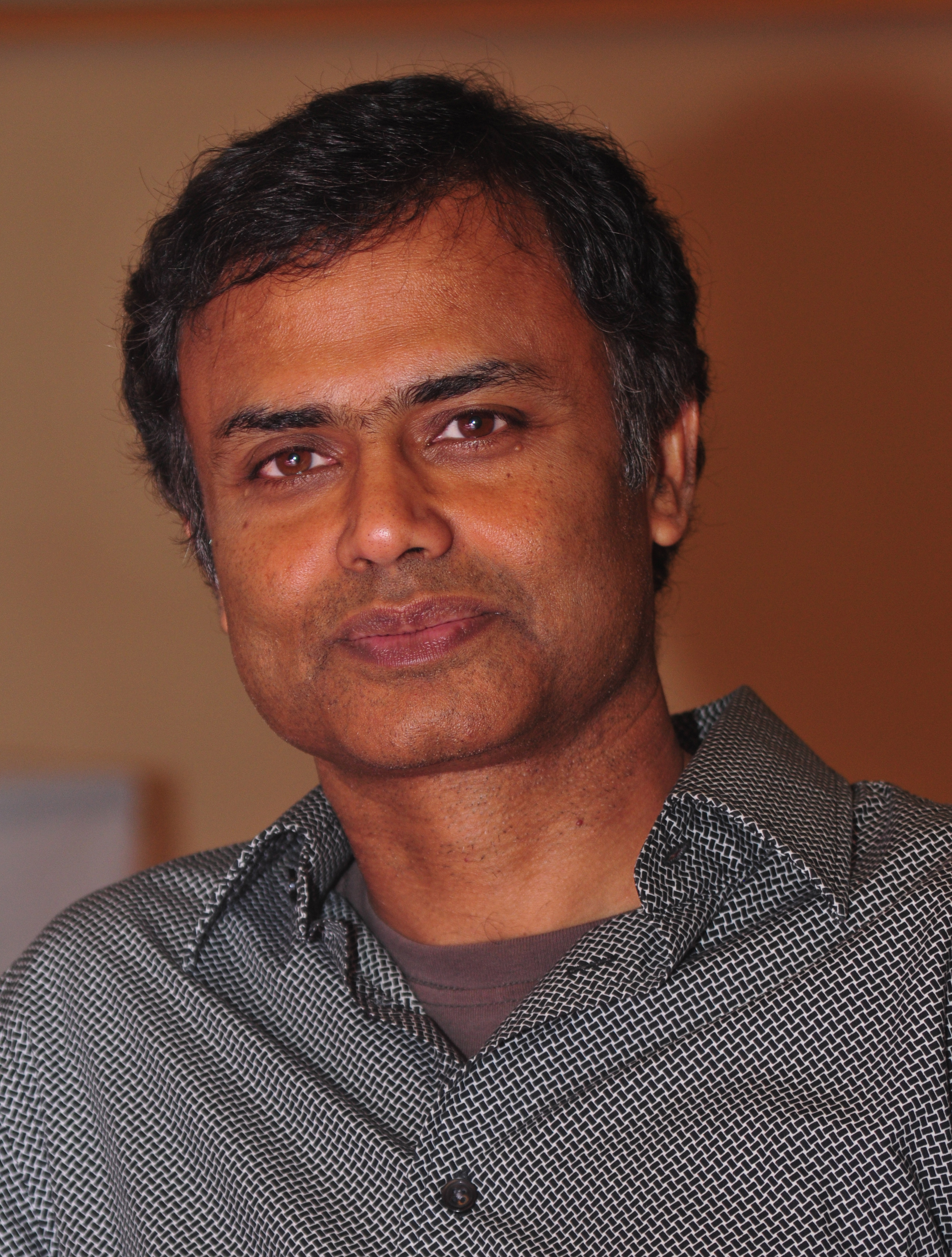}}]{Kaushik Roy} is the Edward G. Tiedemann, Jr., Distinguished Professor of Electrical and Computer Engineering at Purdue University. He received his BTech from Indian Institute of Technology, Kharagpur, PhD from University of Illinois at Urbana-Champaign in 1990 and joined the Semiconductor Process and Design Center of Texas Instruments, Dallas, where he worked for three years on FPGA architecture development and low-power circuit design. His current research focuses on cognitive algorithms, circuits and architecture for energy-efficient neuromorphic computing/ machine learning, and neuro-mimetic devices. Kaushik has supervised more than 100 PhD dissertations and his students are well placed in universities and industry. He is the co-author of two books on Low Power CMOS VLSI Design (John Wiley \& McGraw Hill). 
\end{IEEEbiography}

\vfill

\end{document}